%% file: 0543paper-v2.tex
\documentclass{aa} 
 \usepackage{txfonts}
\usepackage{natbib} \usepackage{graphicx} \usepackage{color}
\graphicspath{{/home/volker/Documents/Figures/AA2008_1/}}
\newcommand{\Lower}[1]{\smash{\lower 1.5ex \hbox{#1}}}
\newcommand\T{\rule{0pt}{2.6ex}}
\newcommand\B{\rule[-1.2ex]{0pt}{0pt}}
\begin{document}

\title{ Cosmic rays and the magnetic field in the nearby starburst
  galaxy NGC\,253 I. The distribution and transport of cosmic rays}

\titlerunning{The distribution and transport of cosmic rays in NGC\,253}

\author{Volker Heesen\inst{1} \and Rainer Beck\inst{2} \and Marita
  Krause\inst{2} \and Ralf-J\"urgen Dettmar\inst{1}}

\institute{Astronomisches Institut der Ruhr-Universit\"at Bochum,
  Universit\"atsstr. 150, 44780 Bochum, Germany \and Max-Planck
  Institut f\"ur Radioastronomie, Auf dem H\"ugel 69, 53121 Bonn,
  Germany}

\date{Received 9 July 2008 / Accepted 28 November 2008}
\abstract {Nearby edge-on
  galaxies showing a synchrotron halo are nearly ideal objects for studying
  the transport of cosmic rays (CRs) in galaxies. Among them, the
  nearby starburst galaxy NGC\,253 hosts a galactic wind indicated by
  various ISM phases in its halo.}
{The diffusive and convective CR transport from the disk into the halo
  is investigated using the local CR bulk speed. The connection between
  the CR transport and the galactic wind is outlined.}
{We observed NGC\,253 with the VLA at $\lambda 6.2\,{\rm cm}$ in a
  mosaic with 15 pointings. The missing zero-spacing flux density of
  the VLA mosaic was filled in using observations with the 100-m
  Effelsberg telescope. We also obtained a new $\lambda 3.6\,{\rm cm}$
  map from Effelsberg observations and reproduced VLA maps at $\lambda
  20\,{\rm cm}$ and $\lambda90\,{\rm cm}$. The high dynamic range
  needed due to the strong nuclear point-like source was addressed
  with a special data calibration scheme for both the single-dish and
  the interferometric observations.}
{We find a thin and a thick radio disk with exponential scaleheights
  of $0.3\,{\rm kpc}$ and $1.7\,{\rm kpc}$ at $\lambda 6.2\,{\rm
    cm}$. The equipartition total magnetic field strength between
  $7\,{\rm\mu G}$ and $18\,{\rm\mu G}$ in the disk is remarkably
  high. We use the spectral aging of the cosmic ray electrons (CREs) seen
  in the vertical profiles of the spectral index to determine a lower
  limit for the global CR bulk
  speed as $170\pm70\,{\rm km\, s^{-1}}$. The linear correlation
  between the scaleheights and the CRE lifetimes, as evident from the
  dumbbell shaped halo, requires a vertical CR
  transport with a bulk speed of $300\pm 30\,{\rm km\, s^{-1}}$ in
  the northeastern halo, similar to the escape velocity of $280\,\rm
  km\,s^{-1}$. This shows the presence of a ``disk wind'' in NGC\,253. In the
  southwestern halo, the transport is mainly diffusive with a diffusion
  coefficient of $2.0\pm0.2\times10^{29}\,\rm cm^2\,s^{-1}$.}
{Measuring the radio synchrotron scaleheight and estimating
  the CRE lifetime allow us to determine the bulk speed of the CR
  transport into the halo. The transport is convective and more
  efficient in the northeastern halo, while it is diffusive in the
  southwestern halo. The luminous material is transported by the
  disk wind, which can explain the different amounts of
  extra-planar \ion{H}{I}, H$\alpha$, and soft X-ray emission in the
  two halo parts. Future low-frequency radio observations will provide
the data to analyze the vertical velocity profile of galactic winds.}
 \maketitle \setcounter{equation}{0} \setcounter{figure}{0}
\setcounter{table}{0} \setcounter{footnote}{0} \setcounter{section}{0}
\setcounter{subsection}{0}
\section{Introduction}
NGC\,253 is a prototypical starburst galaxy at a distance of 3.94\,Mpc
\citep{karachentsev_03a}. It is a late-type spiral galaxy with a high
inclination angle of $78.5^\circ\pm0.5^\circ$ \citep{pence_80a}. This
allows observations of extra-planar components of the interstellar
medium (ISM) at different wavelengths: ROSAT Position Sensitive
Proportional Counter (PSPC) observations have revealed a very extended
halo of soft diffuse X-ray emission extending into the halo to a
distance of $8\,{\rm kpc}$ from the disk. The prominent lobes of soft
X-ray emission show a distinct ``X''-shaped pattern centered on the
nucleus \citep{pietsch_00a}. Moreover, H$\alpha$ observations by
\citet{hoopes_96a} show plumes emerging from the disk into the halo,
which spatially coincide with the lobes of the soft X-ray
emission. Observations of \citet{boomsma_05a} in \ion{H}{I} show large
plumes of extra-planar \ion{H}{I} emission enclosing the soft X-ray
emission.
Given the different phases of the ISM in the halo of NGC\,253, the
question arises of how the observed halo structure was formed. The
distribution of the different ISM phases in shells has led to the
suggestion that the halo can be considered as one huge bubble that is
fueled by an active star-forming region around the nucleus of the
galaxy \citep{strickland_02a}. This scenario is supported by the
finding of a small plume of hot X-ray gas in CHANDRA maps associated
with a conical outflow from the nuclear starburst
\citep{strickland_00a}. Spectroscopic measurements of
H$\alpha$-emitting gas in the southern nuclear outflow cone by
\citet{schulz_92a} give an outflow velocity of $390\,{\rm km\,
  s^{-1}}$. Numerical models by \citet{suchkov_94a} follow the
evolution of a multi-phase halo powered by a compact source located in
the center of the disk. A close resemblance of the ``X''-shaped pattern
of the soft X-ray emission was indeed found, which could be interpreted
as the termination shock of the hot gas where the gas cools
effectively.
A possible explanation for the amount of X-ray and H$\alpha$-emitting
gas in the halo is the dissipation of kinetic energy via shock heating
\citep{lehnert_96a} if the thermalization fraction is high enough
\citep{strickland_02a}. A precondition for this model is the existence
of cool gas in the halo that condenses to clouds from which the shock
can form. ISOPHOT observations discussed by \citet{radovich_01a} show
``X''-shaped far-infrared emission overlapping with the H$\alpha$ and X-ray
emission, indicating the existence of such a cool gas. However, the
origin of the gas is not clear. It could be a preexisting population
of cool gas in the halo, possibly accreted from a merger with a
satellite. Another possibility is that the gas of the disk can be
transported into the halo along the walls of the cavity as
Kelvin-Helmholtz instabilities rip off clumps of cool gas in the disk
\citep[see e.g.][]{suchkov_94a, heckman_00a}. Summarizing, it is
important to understand how the gas is transported from the disk into
the halo.\\
Cosmic rays (CRs) and magnetic fields are important for the evolution
of the ISM, because they may possess an energy density comparable to
that of the thermal gas \citep{beck_96a}. They are even more important
for the study of galactic winds, because the thermal gas itself can
drive a galactic wind only when the heating processes are sufficiently
strong and the gravitational potential is comparably weak. A good
example of this situation is the edge-on galaxy M\,82, which possesses a
huge halo of hot X-ray, H$\alpha$, and radio continuum
emission. However, the hot gas cools radiatively preventing the escape
from the gravitational potential and falls back onto the disk in
condensed clouds. This {\it galactic fountain} scenario proposed by
\citet{field_69a} might explain the infall of the so-called High
Velocity Clouds that could be interpreted as cooled remnants
of the hot gas.
If we incorporate the CR pressure the galactic wind can be sustained
against the radiative losses \citep{ipavich_75a}, because the CRs
suffer only little from radiative losses (mainly via synchrotron
losses of the electrons). As the protons carry most of the energy this
does not influence the CR energy density. Thus the pressure
support from below can drive a galactic wind against the gravity of
the underlying galactic
disk. \citet{breitschwerdt_91a,breitschwerdt_93a} started a series of
papers treating the CR transport along ``flux density tubes'' in a
one-dimensional model. This model was extended to a full
three-dimensional treatment (with rotational symmetry) by
\citet{zirakashvili_96a}, including the influence of galactic
rotation. It was found that CRs can drive a galactic wind under a
broad variety of conditions in the disk and that the loss of mass and
momentum over the lifetime of a galaxy is significant
\citep{ptuskin_97a}. A recent study by \citet{everett_08a} showed that
the extra-planar X-ray emission in the Milky Way can be explained by a
CR driven galactic wind.
The theory of a CR driven galactic wind can be constrained with radio
continuum studies: as the relativistic CRs are the most pervasive
component of the ISM, they are expected to be associated with the
other phases in the halo of NGC\,253. Indeed, a radio halo around
NGC\,253 was detected by \citet{beck_79a} using the 100-m Effelsberg
telescope at $\lambda 3.6\,{\rm cm}$. However, no correction for the
sidelobes of the strong nuclear point-like source was applied and thus it
was not possible to disentangle the intrinsic halo emission from the
contribution by the sidelobes. The study was extended later with VLA
interferometric observations by \citet{carilli_92a}, which show an
extended radio halo over the length of the optical disk very different
in shape from the halo seen in X-rays. The finding of a large ``radio
spur'' southeast of the nucleus already led to speculations of an
outflow connected with the spur.
Extending the observations to other wavelengths and including
polarization, a prediction of the magnetic field structure in the disk
was presented by \citet{beck_94a}. The sensitivity of the
observations, however, was not sufficient to get information for the
extra-planar polarized emission so that the magnetic field structure
in the halo could not be determined. With the new VLA mosaic at
$\lambda 6.2\,{\rm cm}$ we are able to increase the sensitivity at
short wavelengths, allowing us to study in detail the extra-planar
emission and possible relations to the outflow of hot gas. The
advantage is that the high synchrotron losses at this short wavelength
impose tighter constraints on the CR transport from the disk into
the halo. Moreover, for the first time sensitive polarization
measurements were carried out to study the large-scale magnetic
field structure in the halo of NGC\,253, which will be addressed in a
subsequent paper (thereafter Paper~II).
This paper is organized as follows: in Sect.~\ref{sec:cr_observations}
we introduce our observations and present the data reduction. We
present our radio continuum maps along with a discussion of the
morphology in Sect.~\ref{sec:cr_morphology}. The distribution of the
radio spectral index is presented in Sect.~\ref{sec:cr_si}. The
finding of an extended radio halo with a thick radio disk leads to the
determination of the vertical CR bulk speed
(Sect.~\ref{sec:cr_transport}). In Sect.~\ref{sec:cr_discussion} we
discuss the impact of our findings on the feeding of the radio halo
with CRs from the star-forming disk. Finally, we summarize our results
and present our conclusions in Sect.~\ref{sec:cr_conclusions}.
\section{Observations and data reduction}
\label{sec:cr_observations}
\subsection{The Effelsberg observations}
\label{subsec:cr_effelsberg}
For the single-dish observations at $\lambda 3.6\,{\rm cm}$ and
$\lambda 6.2\,{\rm cm}$ of NGC\,253 we used the Effelsberg 100-m
telescope\footnote{The Effelsberg 100-m telescope is operated by the
  Max-Planck Institut f\"ur Radioastronomie (MPIfR).} . The $\lambda
3.6\,{\rm cm}$ observations were obtained between February 2003 and
July 2004 with a single-horn receiver mounted in the secondary focus
cabin (SFK). We used the broad band receiver channel centered at
$8.35\,\rm GHz$ which has a bandwidth of $1.1\,\rm GHz$. The $\lambda
6.2\,{\rm cm}$ observations were carried out in 1997 with a dual-horn
receiver mounted in the SFK using a bandwidth of $500\,\rm MHz$ centered at
$4.85\,{\rm GHz}$. Both receivers are capable to record the total
intensity and the Stokes channels $Q$ and $U$ simultaneously. As
primary (flux) calibrator we used 3C286 adopting the flux density
scale of \citet{baars_77a}.
We observed 26 maps at $\lambda 3.6\,{\rm cm}$ with a map size of
$30\arcmin$ $\times$ $20\arcmin$. Two perpendicular scanning
directions were aligned with the major / minor axis of NGC\,253
($p.a.=52^\circ$), in order to reduce the time for scanning. For the
$\lambda\,{\rm 6.2 cm}$ observations we made use of the ``software
beam switching'' technique described in \citet{morsi_86a}. Since the
observations with the dual-horn are made in the azimuth-elevation
coordinate system, we used a restoration method
\citep[see][]{emerson_79a} to obtain eleven maps with a size of
$30\arcmin$ $\times$ $30\arcmin$.
The data reduction was performed using NOD2 data reduction
package \citep{haslam_74a}. Each map was checked and corrected
for pointing errors using Gaussian fits on the nuclear point-like
source in NGC\,253. The final map was produced using the PLAIT
algorithm, which we used to suppress scanning effects
\citep{emerson_88a}. The noise level of the final map at
$\lambda3.6\,{\rm cm}$ is $500\,\mu\rm{Jy/beam}$ in total power with a
resolution of $84\arcsec$ \footnote{All angular resolutions in this
  paper are referred to as the Half Power Beam Width (HPBW).}. At
$\lambda\,{\rm 6.2 cm}$ a noise level of $1\,\rm{mJy/beam}$ with a
resolution of $144\arcsec$ was achieved.
\begin{figure}[htbp]
\resizebox{\hsize}{!}{ \includegraphics{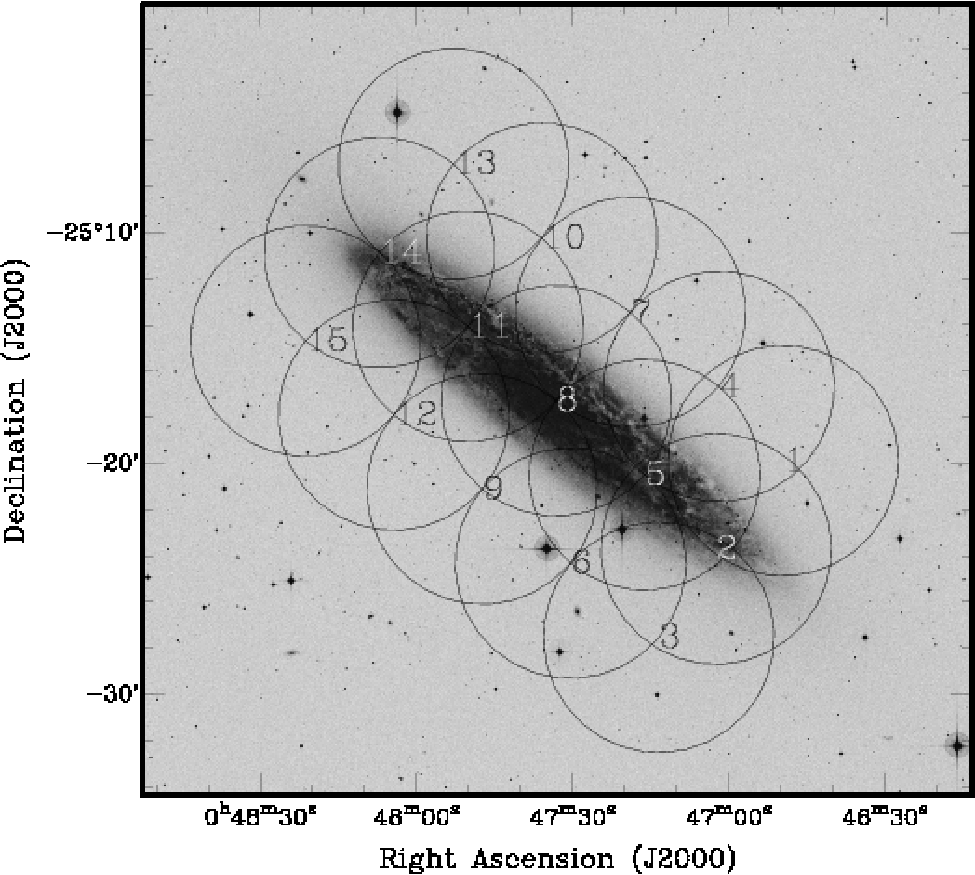}}
\caption[The VLA mosaic at $\lambda 6.2\,{\rm cm}$ of NGC\,253]{The
  VLA mosaic of NGC\,253 at $\lambda 6.2\,{\rm cm}$ consisting of 15
  pointings. The circles indicate the size of the VLA primary beam
  which is $8\arcmin$ at $\lambda 6.2\,{\rm cm}$. The background
  optical image is from the DSS.}
\label{fig:vla_mosaic_62}
\end{figure}
The sidelobes of the Effelsberg beam pattern peak at a level of a few
\% which limits the dynamic range to 100 at most. As the nuclear
point-like source has a flux density of $1\,{\rm Jy}$, we would be
limited to a noise level of $10\,{\rm mJy/beam}$. In order to overcome
these limitations we had to remove the sidelobes of the nuclear
point-like source. This was accomplished by applying a procedure
similar to the CLEAN algorithm by \citet{hoegbom_74a}. This procedure
was revised and extended for application on single-dish observations
by Klein and Rottmann \citep[see][]{mack_95a, rottmann_96a}. The required beam
patterns were obtained from deep observations of 3C84 and 0234-23
at $\lambda 3.6\,{\rm cm}$ and $\lambda 6.2\,{\rm cm}$,
respectively. Each single map of NGC\,253 was cleaned with the mean
parallactic angle of the map, respectively. This is sufficient, since
the variation of the parallactic angle is less than $5^\circ$ during
the observation of a single map.
We were able to obtain essentially noise limited maps with a dynamic
range of at least 1000. The cleaned structures (the difference between
the uncleaned and the cleaned map) exhibit prominent sidelobes
extending far from the disk underlining the importance of a careful
cleaning procedure. Details of the single-dish cleaning can be found
in \citet{heesen_08a}.
%
\subsection{The VLA observations}
\label{subsec:cr_vla}
With the VLA\footnote{The VLA (Very Large Array) is operated by the
  NRAO (National Radio Astronomy Observatory).} in D-configuration we
observed NGC\,253 in July 2004 at $\lambda 6.2\,{\rm cm}$ with a total
integration time of 40~hours. The size of the primary beam of the VLA
antennas at $\lambda 6.2\,{\rm cm}$ is $8\arcmin$, requiring a mosaic
of 15 pointings. They were arranged in three rows parallel to the
major axis of the galaxy, including much of the extra-planar field of
interest. An illustration of the mosaic setup is shown in
Fig.~\ref{fig:vla_mosaic_62} as an overlay onto an optical image from
the DSS\footnote{The compressed files of the "Palomar Observatory -
  Space Telescope Science Institute Digital Sky Survey" of the
  northern sky, based on scans of the Second Palomar Sky Survey are
  \copyright 1993-1995 by the California Institute of Technology and
  are distributed herein by agreement.}.  The data were calibrated
with 3C48 and 3C138 as primary (flux) calibrators using the standard
procedures in the Astronomical Image Processing System\footnote{The
  Astronomical Image Processing System (AIPS) is distributed by the
  National Radio Astronomy Observatory (NRAO) as free software.} with
the flux density scale given in \citet{baars_77a}. The performed
self-calibration allowed us to suppress the sidelobes of the nuclear
point-like source in order to gain a high dynamic range.
After inverting the (u,v)-data and the successive cleaning we
convolved the maps of each pointing with a Gaussian in order to obtain
an identical clean beam. Using LTESS (part of AIPS) the maps were
finally combined with a linear superposition and a correction for the
VLA primary beam attenuation using information from each pointing out
to the 7\,\% level of the primary beam \citep{braun_88a}. Our final
map has a noise level of $30\,{\rm \mu Jy/beam}$ in total power which
corresponds to a dynamic range of $40000$ with respect to the
nucleus.

\begin{figure}[tb]
\resizebox{\hsize}{!}{ \includegraphics{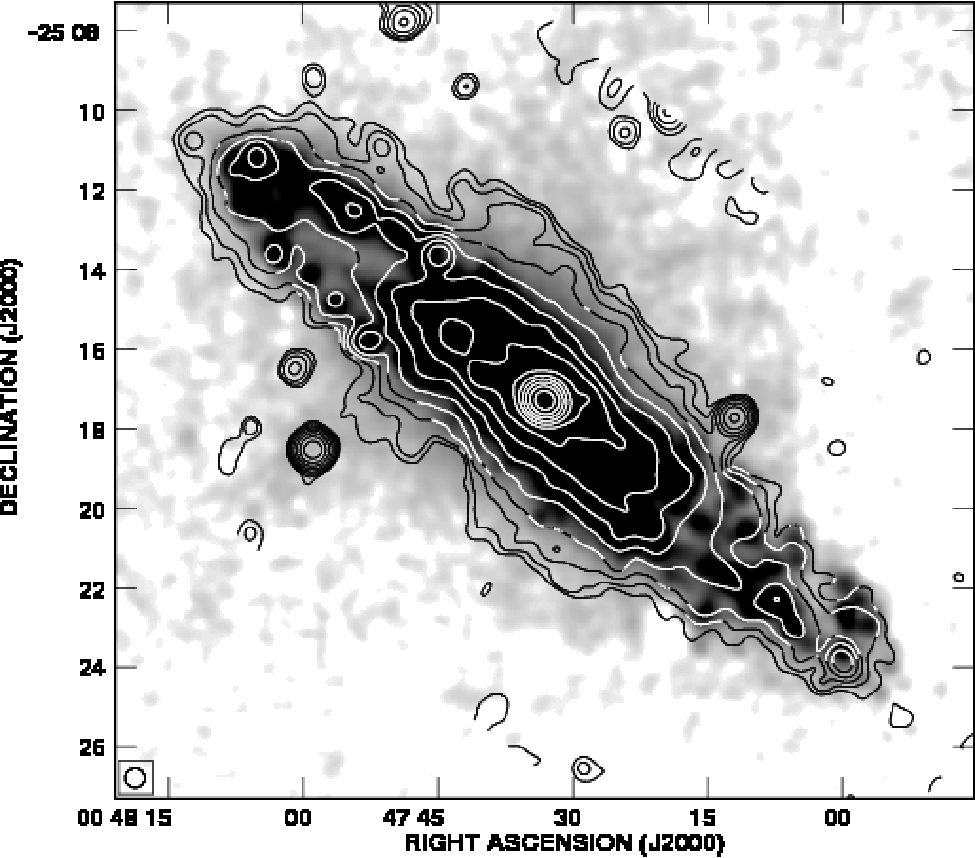}}
\caption{Total power radio continuum at $\lambda 6.2\,{\rm cm}$
  obtained from the VLA observations with $30\arcsec$ resolution. Contours are
  at 3, 6, 12, 24, 48, 96, 192, 384, 768, 1536, 3077, 6144, 12288, and
  24576 $\times$ $30\,{\rm\mu Jy/beam}$. The grey scale shows
  H$\alpha$ emission from \citet{hoopes_96a}.}
\label{fig:n253v6_tp_dss_b30}
\vspace{0.5cm}\resizebox{\hsize}{!}{ \includegraphics{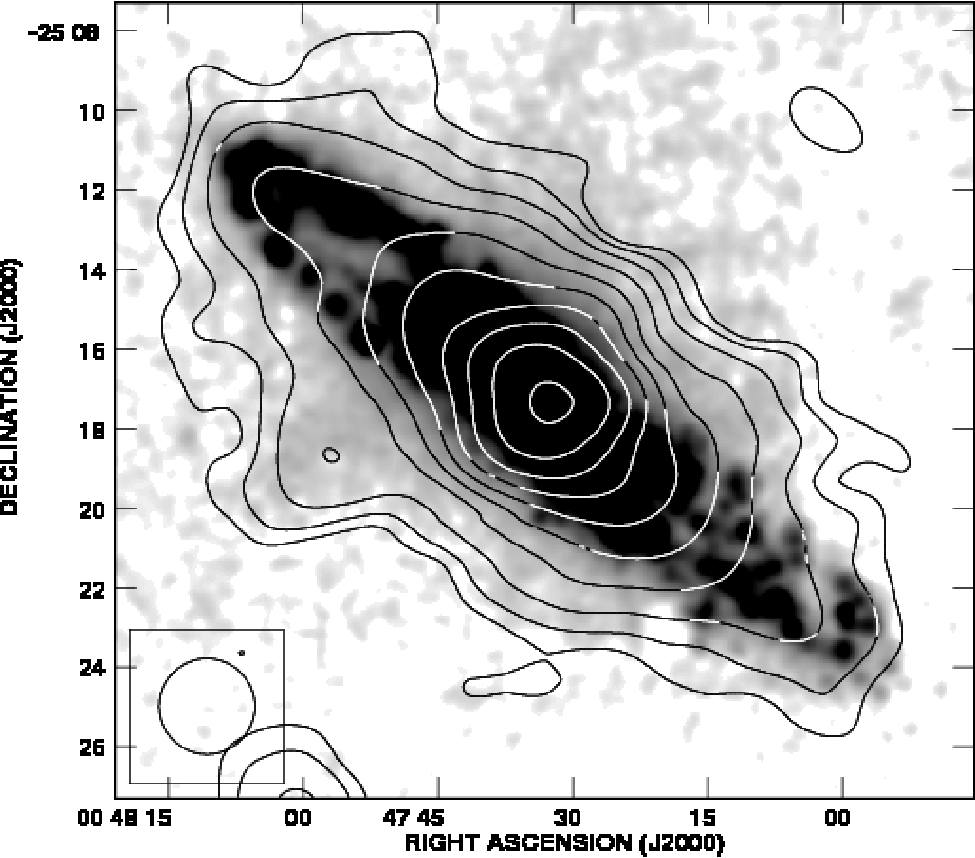}}
\caption{Total power radio continuum at $\lambda 6.2\,{\rm cm}$
  obtained from the Effelsberg observations with $144\arcsec$
  resolution. Contours are at 3, 6, 12, 24, 48, 96, 192, 384, 768, and
  1536 $\times$ $1\,{\rm mJy/beam}$. The grey scale shows H$\alpha$
  emission as in Fig.~\ref{fig:n253v6_tp_dss_b30}.}
\label{fig:n253cm6e_tp_clean_dss_b144}
\end{figure}
\begin{figure}[tb]
\resizebox{\hsize}{!}{ \includegraphics{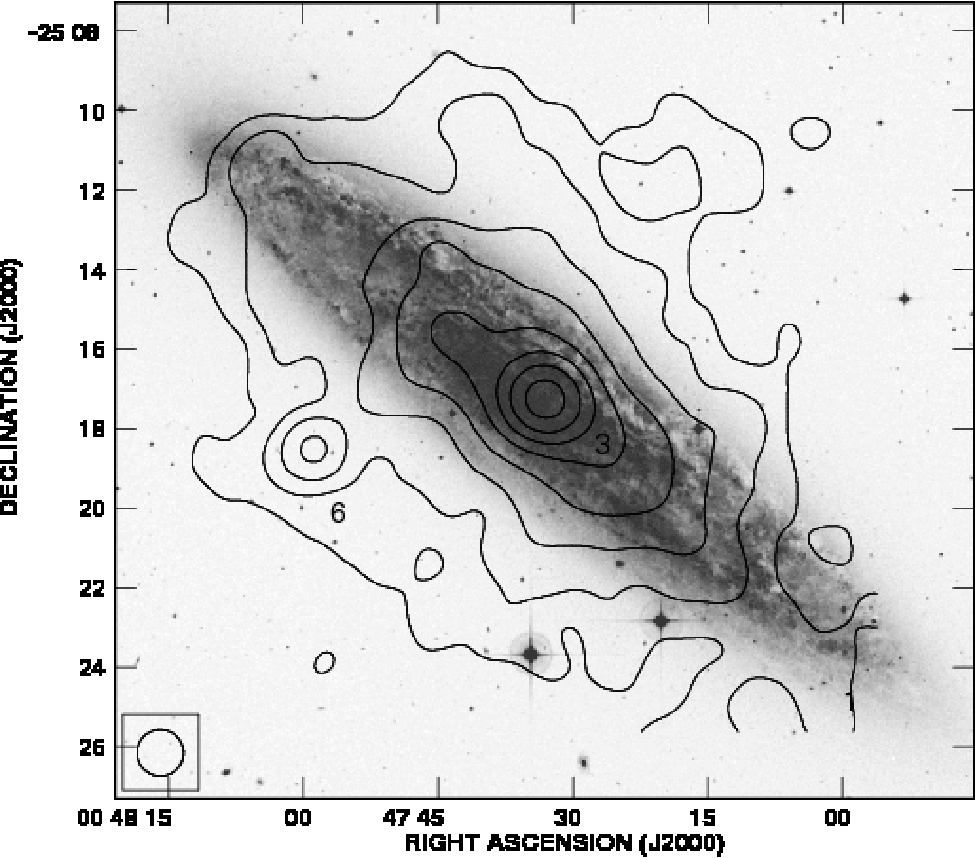}}
\caption[]{Total power radio continuum at $\lambda 90\,{\rm cm}$
  obtained from VLA observations with
  $70\arcsec$ resolution. Contours are at
  3, 6, 12, 24, 48, 96, 192, 384, 768, 1536, and 3077 $\times$
  $8\,{\rm mJy/beam}$. The map is reproduced from \citet{carilli_92a}.}
\label{fig:n253v90_tp_dss_b70}
\setcounter{figure}{05}
\vspace{0.5cm} \resizebox{\hsize}{!}{
  \includegraphics{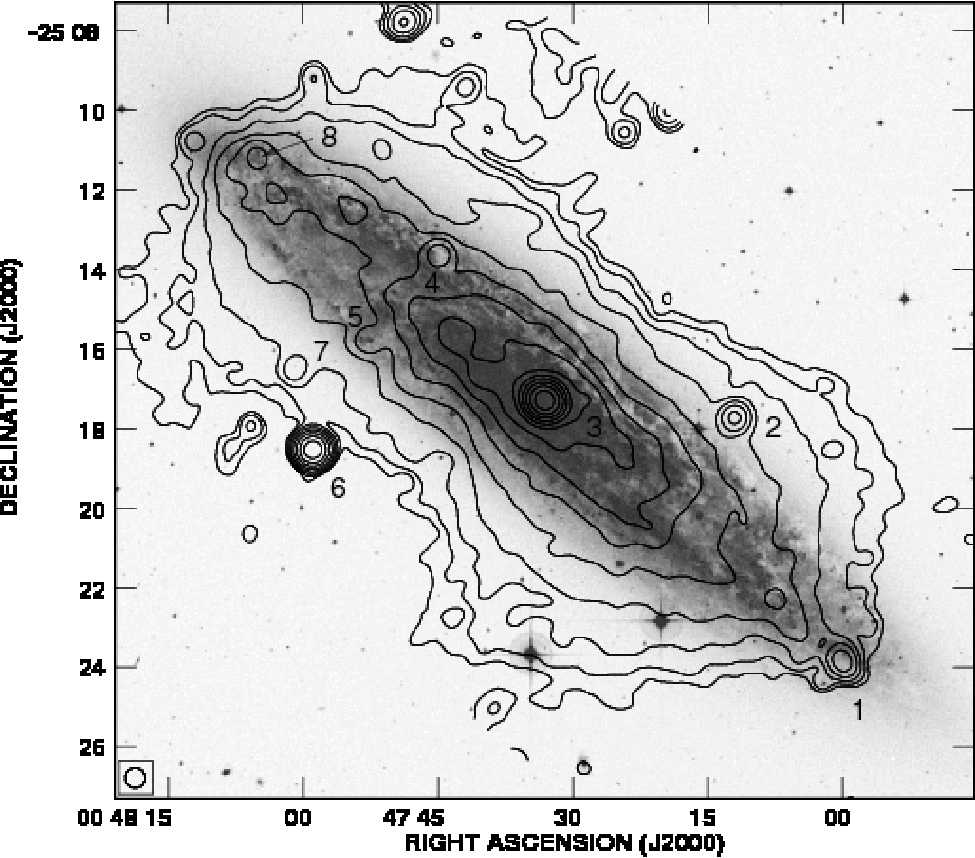}}
\caption[]{Total power radio continuum at $\lambda 6.2\,{\rm cm}$
  obtained from combined VLA and Effelsberg observations with
  $30\arcsec$ resolution. Contours are
  at 3, 6, 12, 24, 48, 96, 192, 384, 768, 1536, 3077, 6144, 12288, and
  24576 $\times$ $30\,{\rm\mu Jy/beam}$.}
\label{fig:n253ve6_tp_dss_b30}
\end{figure}
\begin{figure}[tb]
\setcounter{figure}{04}
\resizebox{\hsize}{!}{ \includegraphics{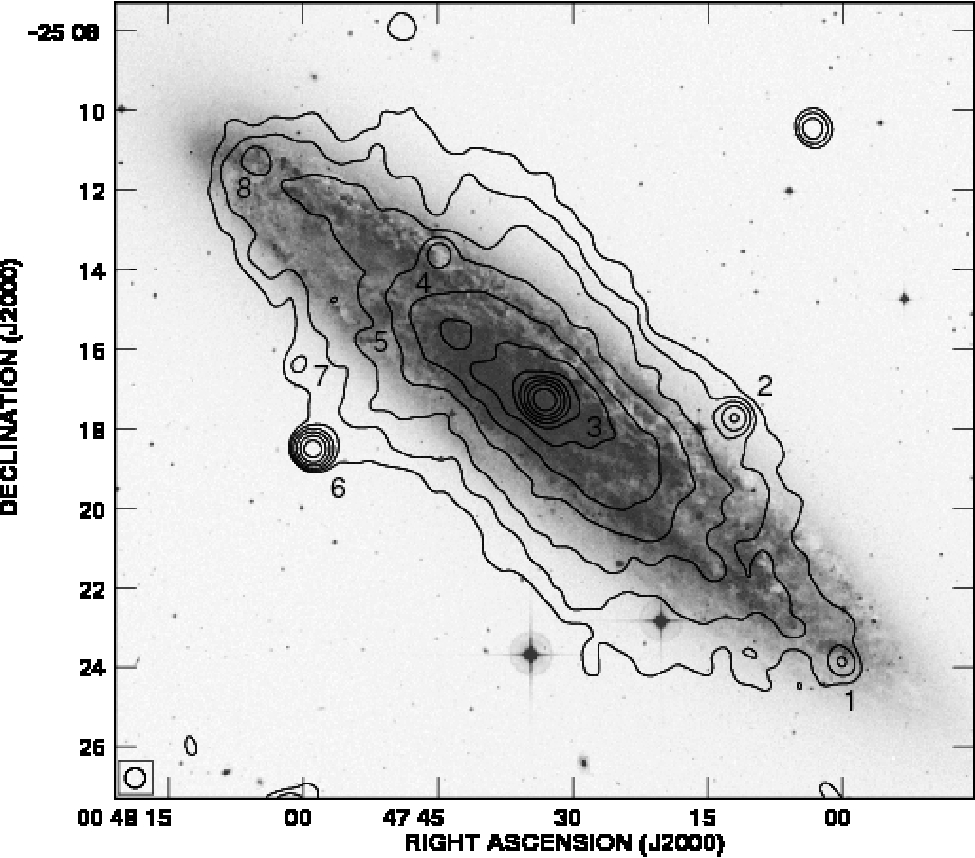}}
\caption[]{Total power radio continuum at $\lambda 20\,{\rm cm}$
  obtained from VLA observations with $30\arcsec$ resolution. Contours
  are at 3, 6, 12, 24, 48,
  96, 192, 384, 768, 1536, and 3077 $\times$ $0.35\,{\rm
    mJy/beam}$. The map is reproduced from \citet{carilli_92a}.}
\label{fig:n253v20_tp_dss_b30}
\setcounter{figure}{06}
\vspace{0.5cm} \resizebox{\hsize}{!}{
  \includegraphics{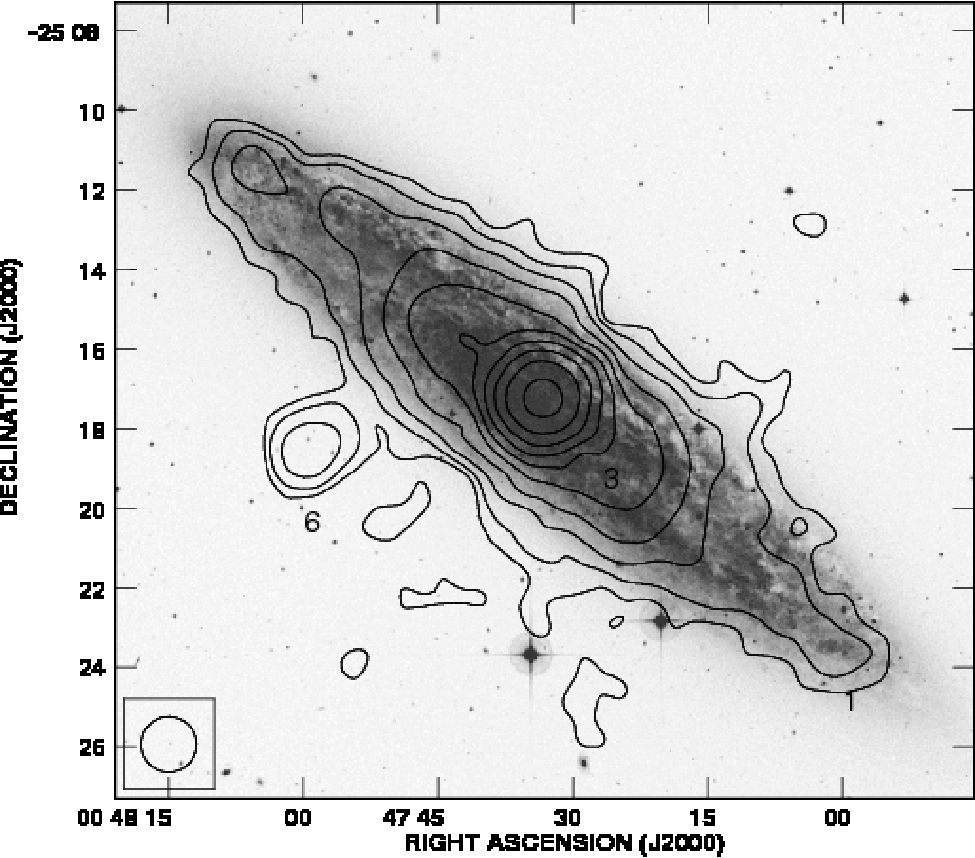}}
\caption[Effelsberg total power radio continuum at $\lambda 3.6\,{\rm
    cm}$.]{Total power radio continuum at $\lambda 3.6\,{\rm cm}$
  obtained from Effelsberg observations with $84\arcsec$
  resolution. Contours are at 3, 6, 12, 24,
  48, 96, 192, 384, 768, and 1536 $\times$ $0.5\,{\rm mJy/beam}$.}
\label{fig:n253cm3e_tp_clean_dss_b84}
\end{figure}
An important shortcoming of interferometric observations is the
limited sensitivity to extended emission on large angular scales. The
VLA in D-configuration at $\lambda 6.2\,\rm cm$ is only sensitive to
emission with a scale smaller than $5\arcmin$. Since NGC\,253 is much
larger, we observe with the VLA only a fraction of the extended
emission detected with a single-dish telescope. The influence of the
missing zero-spacing
flux density can be easily seen by comparing the Effelsberg map at
$\lambda 6.2\,{\rm cm}$ in Fig.~\ref{fig:n253cm6e_tp_clean_dss_b144}
with the VLA map shown in
Fig.~\ref{fig:n253v6_tp_dss_b30}. The missing total flux density is
about 10\,\% of the total flux density measured with Effelsberg; if
the flux density of the extended emission (without the nuclear
point-like source) is considered, the fraction of the missing
zero-spacing flux density increases to 20\,\%.
We used the Effelsberg map in order to fill in the missing
zero-spacing flux in the VLA map. This was done using IMERG (part of
AIPS) which performs a Fourier combination of the two maps. We stress
that in particular for mosaics, which cover the entire region of
 interest, a single-dish map is necessary in order to detect all
extended emission.
The $\lambda 90\,{\rm cm}$ and $\lambda 20\,{\rm cm}$ maps are from
VLA observations by \citet{carilli_92a}. While one pointing with a
primary beam size of $138\arcmin$ is sufficient to cover NGC\,253 at
$\lambda 90\,{\rm cm}$, three pointings with a beam size of
$25\arcmin$ were combined to a mosaic at $\lambda 20\,\rm cm$. The
largest angular scale structure which can be imaged reasonable well
with the VLA is $70\arcmin$ at $\lambda 90\,\rm cm$ and $15\arcmin$ at
$\lambda 20\,\rm cm$, respectively. The needed correction for the
missing zero-spacing flux of the mosaic at $\lambda 20\,\rm cm$ was
not performed because no single-dish map is available at this
wavelength. Hence, the $\lambda 20\,\rm cm$ map is not well suited to
investigate the structure of the extended emission in the halo.
%
\section{Morphology of the total power emission}
\label{sec:cr_morphology}
We present the total power maps as contour plots with the first
contour indicating emission at three times the rms noise level
($3\,{\rm rms}\times2^{m-1}$; $m=1,2,3,\ldots$). The optical background
image was obtained from the DSS. A summary of the maps presented
in this Paper can be found in Table~\ref{tab:cr_overview_observations}.
\begin{table}[tbhp]
\caption{The maps of NGC\,253 included in this paper.}
\begin{center}
\begin{tabular}{llllll}
\hline\hline
$\lambda$ \T & resolution & rms & dynamic & references\\
$[\rm cm]$ \B & & $[\rm mJy/beam]$ & range &\\\hline
$90$ \T & $70\arcsec$ & $8$ & 500 & \cite{carilli_92a}\\
$20$ & $30\arcsec$ & $0.35$ & 5700 & \cite{carilli_92a}\\
$6.2$ & $30\arcsec$ & $0.03$ & 42000 & this paper\\
$3.6$ \B & $84\arcsec$ & $0.5$ & 2000 & this paper\\\hline
\end{tabular}
\end{center}
\label{tab:cr_overview_observations}
\end{table}
\begin{table*}[bthp]
\caption[]{Unresolved sources which are denoted in Figs.\ \ref{fig:n253v90_tp_dss_b70}-\ref{fig:n253cm3e_tp_clean_dss_b84}. The
  error of the flux densities was calculated assuming a 5\,\%
  calibration error.}
\begin{center}
\begin{tabular}{l|lllllll}
\hline\hline No. \T\B & RA & Dec & $S_{\lambda90}~{\rm [mJy]}$ &
$S_{\lambda20}~{\rm [mJy]}$ & $S_{\lambda6.2}~{\rm [mJy]}$ &
$S_{\lambda3.6}~{\rm [mJy]}$ & $\alpha$\\\hline
1 \T & $00^{\rm h} 47^{\rm m} 00^{\rm s}.02$ & $-25^\circ 23' 50''.1$ & &
$4.1\pm 0.2$ & $2.2\pm 0.1$ & & $0.5\pm0.2$\\ 
2 & $00^{\rm h} 47^{\rm m} 12^{\rm s}.01$ & $-25^\circ 17' 43''.9$ & &
$17.3\pm 1.0$ & $6.6\pm 0.3$ & & $0.8\pm0.2$\\ 
3 & $00^{\rm h} 47^{\rm m} 33^{\rm s}.12$ & $-25^\circ 17' 17''.3$ &
$3960\pm 80$ & $2040\pm 100$ & $1270\pm 60$ & $980\pm 50$ &
$0.44\pm0.02$\\ 
4 & $00^{\rm h} 47^{\rm m} 44^{\rm s}.91$ & $-25^\circ
13' 38''.4$ & & $26\pm 2$ & $7.9\pm 0.4$ & & $1.0\pm0.2$\\ 
5 & $00^{\rm h} 47^{\rm m} 52^{\rm s}.41$ & $-25^\circ
15' 44''.8$ & & & $1.5\pm 0.2$ & &\\ 
6 & $00^{\rm h} 47^{\rm m} 58^{\rm s}.90$ & $-25^\circ 18' 31''.9$ &
$190\pm 10$ & $53\pm 3$ & $19\pm 1$ & $9.5\pm 0.5$ & $0.84\pm0.02$\\ 
7 & $00^{\rm h} 48^{\rm m} 00^{\rm s}.84$ & $-25^\circ 16' 26''.3$ & &
$2.1\pm 0.3$ & $1.10\pm 0.06$ & & $0.5\pm0.2$\\ 
8 \B & $00^{\rm h} 48^{\rm m} 05^{\rm s}.0$ & $-25^\circ 11' 11''.8$ & &
$5.3\pm 0.3$ & $3.05\pm 0.15$ & & $0.4\pm0.2$\\\hline
\end{tabular}
\end{center}
\label{tab:cr_point_sources}
\end{table*}
The continuum emission not only extends over the optical disk
along the major axis, but we also find a notable extension of emission
along the minor axis in all maps. We refer to this as {\it
  extra-planar} emission. A detailed analysis and proof that the
emission does not belong to the disk is given in
Sect.~\ref{subsec:cr_the_thick_radio_disk}. The extent of extra-planar
emission is strongly dependent on the observation wavelength, which can
be seen by comparing the maps. The $\lambda 90\,{\rm cm}$
map shown in Fig.~\ref{fig:n253v90_tp_dss_b70} has very pronounced
extra-planar emission, while the disk does not show up prominently.
The $\lambda 20\,\rm cm$ and $\lambda 6.2\,\rm cm$ maps
presented in Figs.\ \ref{fig:n253v20_tp_dss_b30} and
\ref{fig:n253ve6_tp_dss_b30} have less extra-planar emission and at
$\lambda 3.6\,\rm cm$ there is almost no extra-planar emission
(Fig.~\ref{fig:n253cm3e_tp_clean_dss_b84}).
The morphology of the total power emission also depends on the
resolution and the sensitivity of the maps. In the $\lambda 20\,{\rm
  cm}$ and $\lambda 6.2\,{\rm cm}$ maps with a resolution of
$30\arcsec$ the spiral arms are distincted features in the disk, one
extending to the northeast and another to the southwest. The inner
part of the disk is visible as a plateau of high emission coinciding
with the brightest part in the optical image. In all maps, independent
of the resolution, there is an extension southeast of the nucleus with
an unresolved source at its southern tip. This is an example of a
so-called ``radio spur'', where the magnetic field has a strong
vertical component, as will be shown in Paper~II.
The total extent of emission perpendicular to the disk is not a good
measure for the amount of extra-planar emission since it depends on
the sensitivity of the observations. This is visible in the $\lambda
20\,{\rm cm}$ map which shows less extra-planar emission than the
$\lambda 6.2\,{\rm cm}$ map because its noise level is 10 times
higher. Thus, with a low sensitivity we observe only the brightest tip
of the radio halo. Moreover, the extent appears to be larger for low
resolutions. This is the motivation to use the scaleheight $h$ which
prescribes the emission profiles perpendicular to the disk as an
exponential function $\propto \exp(-z/h)$ of the distance $z$ from the
galactic midplane. In Sect.~\ref{subsec:cr_the_thick_radio_disk} we
discuss the scaleheights in order to characterize the radio halo.
We find that the extent of the extra-planar emission varies with
the offset along the major axis. The contour lines in all maps show a dumbbell
shape which means that the extension of the extra-planar emission has
a local minimum at the center of the disk. While this trend is visible in
all maps it is clearly seen at the shorter
wavelengths. Especially the $\lambda 3.6\,{\rm cm}$ map
has a prominent neck above and below the nucleus, where the contour
lines indicate a steep gradient along the minor axis. This points to a
smaller scaleheight of the extra-planar emission. If the scaleheight
were constant along the major axis, the largest vertical extension of
the emission would be expected in the center. We investigate this
issue and give an explanation in Sect.~\ref{sec:cr_transport}.
\section{Spectral index distribution}
\label{sec:cr_si}
\subsection{Unresolved sources}
\label{subsec:cr_si_points}
Our continuum maps, in particular those with the highest resolution,
show several unresolved (point-like) sources. We identified the
sources in order to subtract them from the maps prior to our analyis
of the extended emission. We obtained peak flux
densities and positions by fitting Gaussians to them, where we
corrected the peak flux density for the underlying extended
emission. This is particularly important for the nuclear point-like
source, because it is embedded in strong extended emission. The list of
point-like sources presented in Table~\ref{tab:cr_point_sources}
includes positions, flux densities, and spectral indices. The spectral $\alpha$
is defined by the relation $S\propto\nu^{-\alpha}$, where $S$ is the
total power radio continuum flux density and $\nu$ is the observation
frequency. The strongest two sources,
the nucleus (no.\ 3) and the source in the radio spur (no.\ 6), are
detected at all four wavelengths.  For these two sources the spectral
index and its error is calculated in the range between $\lambda
90\,{\rm cm}$ and $\lambda 3.6\,{\rm cm}$. Point source no.\ 6 was
resolved with VLA observations in B-configuration into two components with
$2\arcsec$ separation \citep{carilli_92a}. This source that has
no known optical counterpart is likely a background AGN.
\subsection{Total flux density}
\label{subsec:cr_si_total}
The total radio continuum flux density was integrated within ellipses
with the major axis aligned to the position angle ($p.a.=52^\circ$) of
NGC\,253. The ellipses were chosen such that the visible extended
  emission in the disk and halo is covered. The parameters of the
flux density integration are listed in Table~\ref{tab:cr_ellipses} and
the total flux densities obtained at different measurements are listed
in Table~\ref{tab:cr_tp_flux density}. The flux densities of the
point-like sources were subtracted from the total flux density, in
order to obtain the flux density of the extended emission only.
\begin{figure}[tb]
\resizebox{\hsize}{!}{ \includegraphics{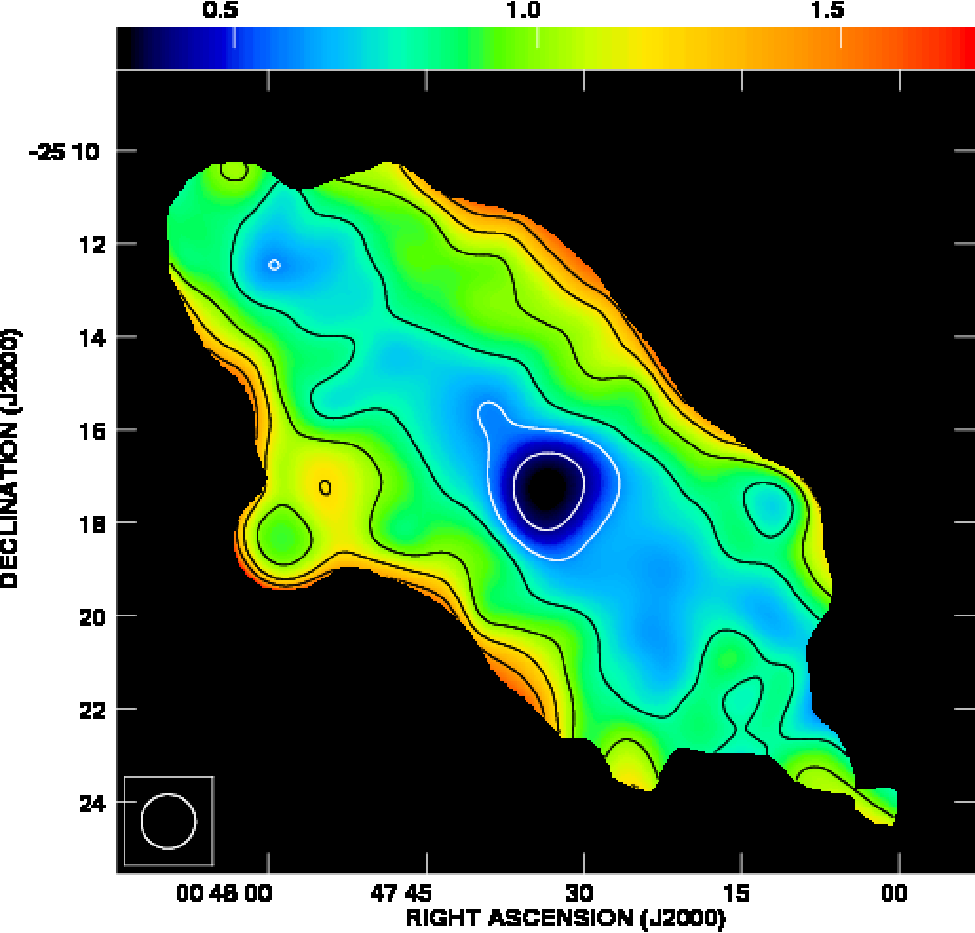}}
\caption{Spectral index
  distribution between $\lambda 90\,{\rm cm}$ and $\lambda 20\,{\rm
    cm}$ with $70\arcsec$ resolution. Contours are at 0.6, 0.8, 1.0,
  1.2, and 1.4. The total power radio continuum maps were clipped at a
  level of $4\times$ the rms noise prior to the combination, resulting
  in a spectral index error of less than $\pm0.3$.}
\label{fig:n253_spix_90_20}
\vspace{0.5cm}\resizebox{\hsize}{!}{ \includegraphics{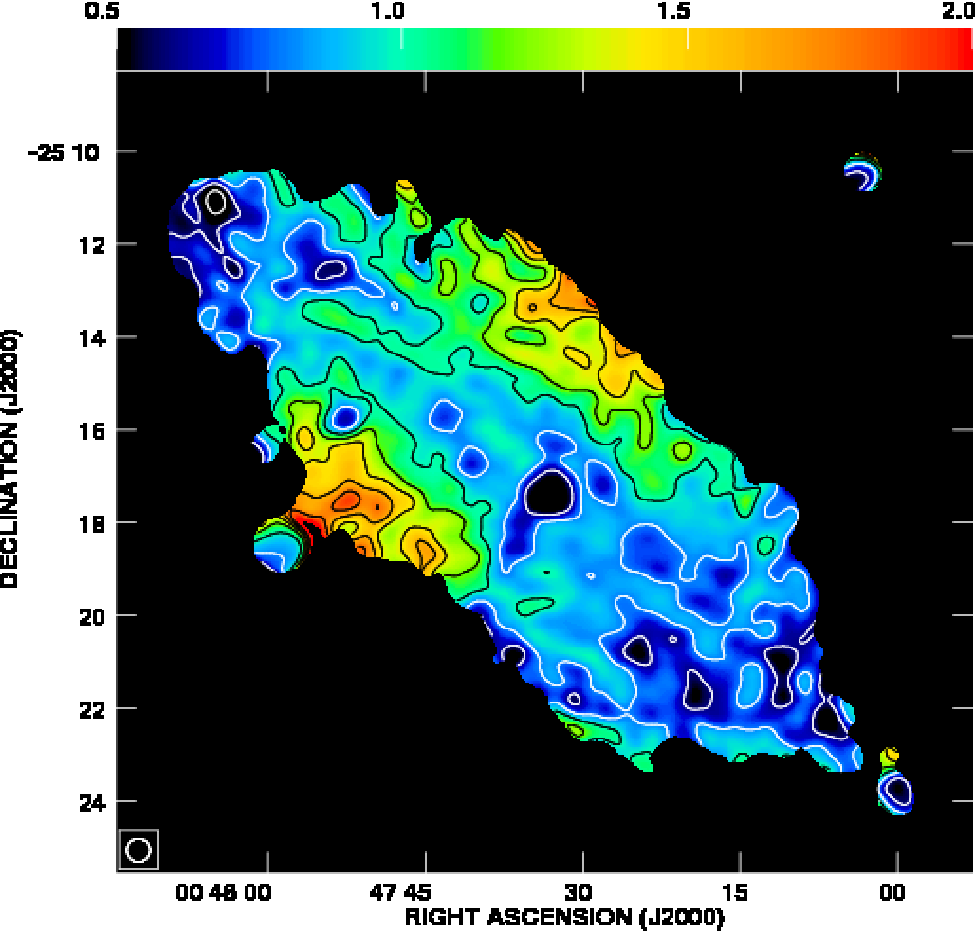}}
\caption{Spectral index
  distribution between $\lambda 20\,\rm cm$ and $\lambda 6.2\,{\rm cm}$ with
  $30\arcsec$ resolution. Contours are at 0.6, 0.8, 1.0, 1.2, 1.4,
  1.6, and 1.8. Both total power radio continuum maps were clipped at
  a level of $5\times$ the rms noise prior to the combination,
  resulting in a spectral index error of less than $\pm0.4$.}
\label{fig:n253_spix_20_6}
\end{figure}
\begin{figure}[htb]
\resizebox{\hsize}{!}{ \includegraphics{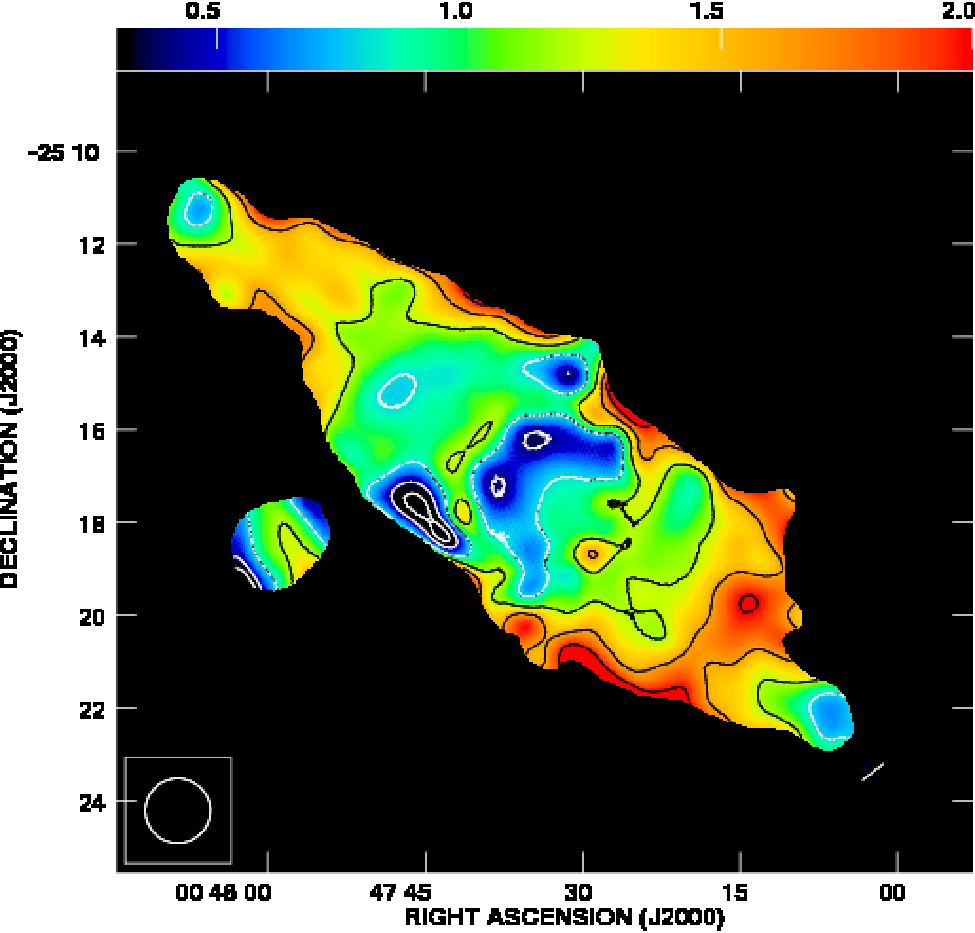}}
\caption{Spectral index distribution between $\lambda 6.2\,\rm cm$ and
  $\lambda 3.6\,{\rm cm}$ with $84\arcsec$ resolution. Contours are at
  0.2, 0.4, 0.8, 1.2, 1.6, and 2.0. Both total power radio continuum
  maps were clipped at a level of $7\times$ the rms noise prior to the
  combination, resulting in a spectral index error of less than
  $\pm0.5$.}
\label{fig:n253_spix_90_3}
\end{figure}
The average spectral index of the extended emission is $\alpha_{\rm
  e}=0.90\pm 0.08$ in the range between $\lambda 90\,{\rm cm}$ and
$\lambda 3.6\,{\rm cm}$. This spectrum is considerably steeper than that
of young cosmic ray electrons (CREs) in supernova remnants (SNRs)
with $\alpha\approx0.5$, although young shell SNRs can have a steeper
spectrum with $\alpha\approx0.7$ \citep[see e.g.][]{green_02a}. This
is an indication for synchrotron losses of the CREs, which is be
discussed in
Sect.~\ref{subsec:loss_processes_of_the_cosmic_ray_electrons}.
Assuming a thermal fraction of 8\,\% at $\lambda 20\,\rm cm$
\citep{niklas_97a}, we find a synchrotron non-thermal spectral index
of $\alpha_{\rm nt}=1.0$.
\subsection{Extended emission}
\label{subsec:cr_si_extended}
\begin{table}[tbh]
\caption[]{The total flux densities were integrated within ellipses
  chosen to include the extended emission in the disk and halo. The
  major axis $a$, and the minor axis $b$ for each wavelength $\lambda$ is
  presented.}
\begin{center}
\begin{tabular}{lllll}
\hline\hline $\lambda~{\rm [cm]}$ \T\B & $a~{\rm [arcsec]}$ & $a~{\rm
  [kpc]}$ & $b~{\rm [arcsec]}$ & $b~{\rm [kpc]}$\\\hline
90 \T & 620 & 12.1 & 392 & 7.6\\ 
20 & 620 & 12.1 & 392 & 7.6\\ 
6.2 & 680 & 13.3 & 430 & 8.4\\ 
3.6 \B & 648 & 12.6 & 400 & 7.8\\\hline
\end{tabular}
\end{center}
\label{tab:cr_ellipses}
\end{table}
In this section we present the spectral index distribution of the
extended emission, which we obtained from the total power maps. The
center of all spectral index maps is dominated by the nuclear
point-like source (no.\ 3) with its flat spectral index of
$\alpha=0.44$. In Fig.~\ref{fig:n253_spix_90_20} we present the
$\lambda 90\,{\rm cm}/\lambda 20\,{\rm cm}$ spectral index map. The
$\lambda 20\,{\rm cm}$ map was convolved with a Gaussian to match the
resolution of the $\lambda 90\,{\rm cm}$ map with $70\arcsec$. At this
low resolution the inner disk and the spiral arms are not resolved. We
find that the spectral index gradually steepens with increasing
distance from the major axis, with the gradient strongest in the inner
part of the disk. There is a slight trend that indicates that the
southwestern part possesses a flatter spectral index than the
northeastern part. At the southern tip of the radio spur the
point-like source no.\ 6 is recognized by its flatter spectral index of
$\alpha=0.84$ compared with the spectral index in the halo.
The $\lambda 20\,{\rm cm}/\lambda 6.2\,{\rm cm}$ spectral index map
presented in Fig.~\ref{fig:n253_spix_20_6} with its higher resolution
of $30\arcsec$ reveals more details. The spiral arms are well resolved
and show a flatter spectral index than the inter-arm regions. In
addition the inner disk has a notable plateau with a flat spectral
index of $6\arcmin$ extent along the major axis, which is not so
clearly seen in the other maps. We find regions of steep spectral
index $(\alpha>1)$ at the base of the radio spur and at the
northeastern extension. This finding agrees with the $\lambda 90\,{\rm
  cm}/\lambda 20\,{\rm cm}$ spectral index map presented above, which
has a steep spectral index at these locations, too.
The $\lambda 90\,{\rm cm}/\lambda 3.6\,{\rm cm}$ spectral index map
presented in Fig.~\ref{fig:n253_spix_90_3} has the lowest angular
resolution of $84\arcsec$ among our spectral index maps. The
morphology of this map is similar to the $\lambda 20\,{\rm cm}/\lambda
6.2\,{\rm cm}$ map even though smoothed out in comparison. The
spectral index is flat in the inner disk and in the spiral arms with
$\alpha<0.7$. Limited by the sensitivity of the $\lambda 3.6\,{\rm
  cm}$ map we find only four small extensions with extra-planar
spectral indices, which are steeper than in the disk. The exception is
the extension northwest of the nucleus, which contains the point-like
source no.\ 2 with a flat spectral index of $\alpha=0.68$.
\begin{table}[bth]
\caption[]{Integrated total power flux densities. $S_{\rm i}$ is
  the total integrated flux density and $S_{\rm e}$ is the flux
  density of the extended emission only. The error of the flux
  densities was calculated as the quadratic sum of a 5\,\% calibration error and the baselevel error.}
\begin{center}
\begin{tabular}{lllll}
\hline\hline $\lambda~{\rm [cm]}$ \T\B& Instrument & $S_{\rm i}~{\rm
  [Jy]}$ & $S_{\rm e}~{\rm [Jy]}$\\\hline
90 \T & VLA & $16.5\pm 1.9$ & $12.5\pm 1.9$\\ 
20 & VLA & $6.3\pm 1.1$ & $4.2\pm 1.1$\\ 
6.2 & VLA + Effelsberg & $2.71\pm 0.14$ & $1.44\pm 0.07$\\ 
3.6 \B & Effelsberg & $1.66\pm 0.08$ & $0.68\pm0.04$\\\hline
 \end{tabular}
\end{center}
\label{tab:cr_tp_flux density}
\end{table}
\section{Cosmic ray transport}
\label{sec:cr_transport}
While the CRs are created and accelerated in the disk, the observation
of extra-planar radio continuum emission requires a transport of CRs
from the disk into the halo. This transport is investigated in
this section.
\subsection{The scaleheight of the thick radio disk}
\label{subsec:cr_the_thick_radio_disk}
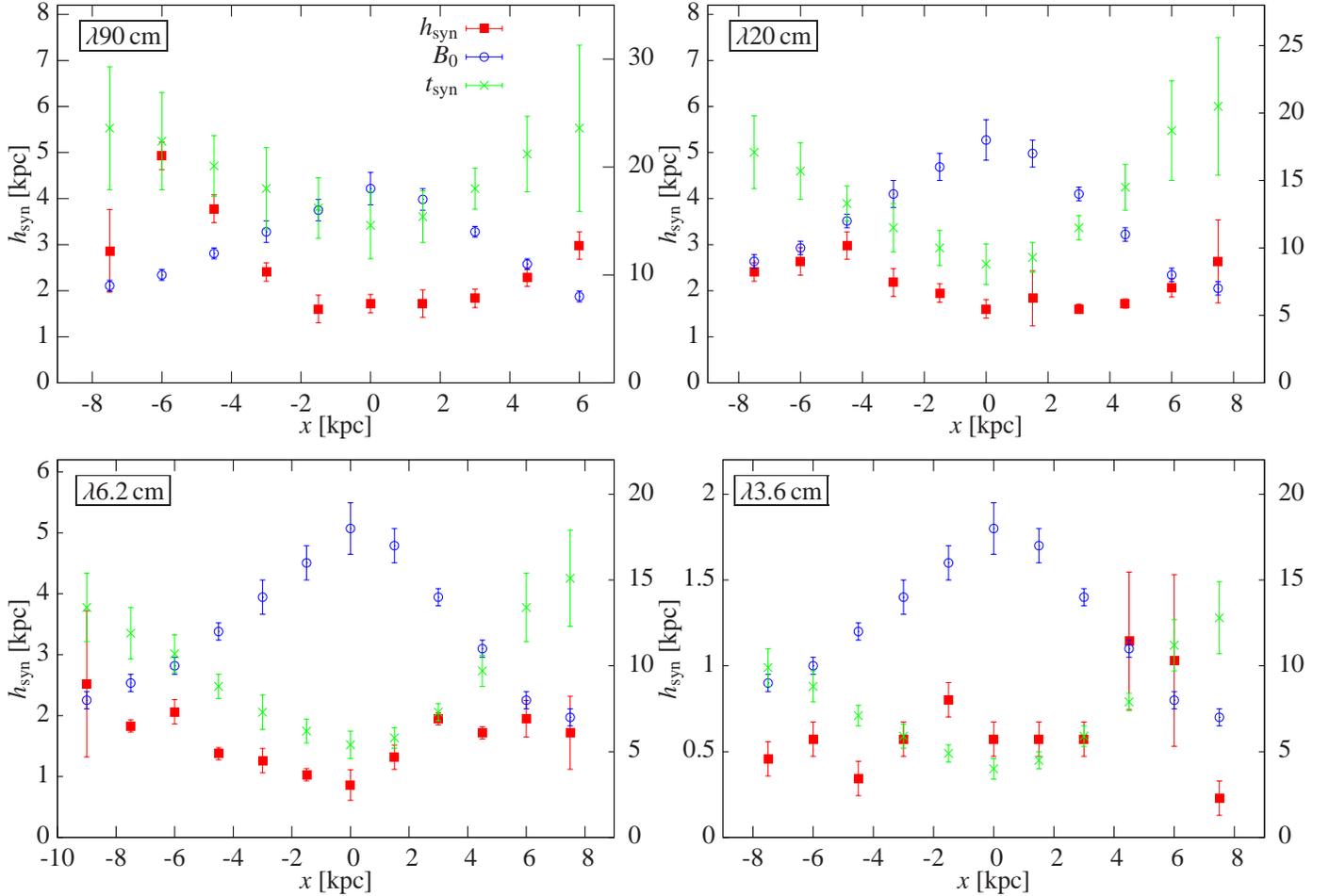
\begin{figure*}[tb]
\resizebox{\hsize}{!}{
  \Large\input{0543fig11.tex}}
\caption{Scaleheight $h_{\rm syn}$ as a function of the offset $x$
  along the major axis (northeast is positive and southwest is
  negative). The magnetic field strength $B_0$ (in units of $\mu\rm G$
  on the right axis) and the electron lifetime $t_{\rm e}$ (in units
  of $10^6\,\rm yr$ on the right axis) are also shown.}
\label{fig:cr_scaleheights_90}
\end{figure*}

With the observed inclination angle of $i=78.5^\circ\pm0.5^\circ$
\citep{pence_80a} any emission profile 
taken perpendicular to the disk is a superposition of disk and halo
emission. We created a template of the vertical emission profile by
projecting the emission profile along the major axis onto the minor
axis. Moreover, the profile was
convolved to the resolution of the observations in order to account
for the beam smearing. Thus, the {\it effective beam size} was
determined by fitting a Gaussian to the resulting profile. Any
emission profile perpendicular to the disk that does not exceed the
effective beam size cannot be attributed to extra-planar emission but
is just the observed inclined emission of the disk. Hence, we arrived
at effective beam sizes of $1.2\arcmin$, $1.2\arcmin$, $2.1\arcmin$,
and $2.5\arcmin$ for $\lambda90\,{\rm cm}$, $\lambda20\,{\rm cm}$,
$\lambda6.2\,{\rm cm}$, and $\lambda3.6\,{\rm cm}$, respectively.
For the following analysis of the extended emission we subtracted all
point-like sources by fitting Gaussians to them. The emission profiles
were obtained using a strip integration perpendicular to the major
axis with a strip width of $1.5\arcmin$ ($1.7\,{\rm kpc}$) and a pixel
separation of $0.25\arcmin$ ($0.29\,{\rm kpc}$). With the routine
discussed in \citet{dumke_95a} profiles
consisting of two exponential functions were fitted to the measured
profiles taking the effective beam size into account. This yields a
scaleheight of the thin and the thick radio disks for the northern and
the southern hemisphere, respectively. The synchrotron scaleheight
$h_{\rm syn}$ is calculated as the mean of the scaleheight of the
thick radio disk in the northern and the southern hemisphere. In case
we could not obtain a two-component fit, we took the one-component
exponential scaleheight as the thick radio disk.
The scaleheights are smallest in the center and increase along the
major axis further away from the center. This is shown in
Fig.\,\ref{fig:cr_scaleheights_90} where also the variations of the
total magnetic field strength and the CRE lifetime are presented. The
error of the scaleheights is estimated by varying the choice of data
points. Moreover, the difference between the northern and the southern
scaleheights contributes to the error.  The weighted mean scaleheight
at $\lambda 6.2\,\rm$ of the thin radio disk is
$0.33\arcmin\pm0.05\arcmin$ ($0.38\pm0.06\,\rm kpc$). The thick radio
disk at $\lambda 6.2\,\rm cm$ has a scaleheight of
$1.5\arcmin\pm0.1\arcmin$ ($1.7\pm0.1\,{\rm kpc}$). At $\lambda20\,\rm
cm$ scaleheights of $0.38\arcmin\pm0.05\arcmin$ ($0.44\pm0.06\,{\rm
  kpc}$) for the thin radio disk and $1.5\arcmin\pm0.1\arcmin$
($1.7\pm0.1\,{\rm kpc}$) for the thick radio disk are found. Note that
the $\lambda 20\,\rm cm$ profiles drop beyond $3\arcmin$ height above
the disk due to missing large-scale structure. However, this has only
a marginal effect on the scaleheights.
The $\lambda6.2\,{\rm cm}$ and $\lambda20\,{\rm cm}$ maps have a
sufficient signal-to-noise (S/N) ratio, so that we were able to obtain
two-component exponential distributions, whereas the $\lambda3.6\,{\rm
  cm}$ map does not have a sufficient S/N ratio so that we used only a
one-component exponential distributions. At $\lambda3.6\,\rm cm$ we
found a mean scaleheight of $0.5\arcmin\pm0.1\arcmin$ ($0.6\pm0.1\,\rm
kpc$) and at $\lambda 90\,\rm cm$ we found $2.2\arcmin\pm0.2\arcmin$
($2.5\pm0.2\,\rm kpc$). The scaleheights at $\lambda 3.6\,\rm cm$ are
dominated by the thin disk, because the low signal does not allow the
thick disk to be traced to sufficiently large distances. In contrast at
$\lambda 90\,\rm cm$ the emission is dominated by the thick radio disk
so that one-component exponential fits are sufficient.
The average scaleheight of NGC\,253 agrees well with those reported
for other edge-on galaxies at $\lambda 6\,\rm cm$
(see also Sect.~\ref{sec:cr_discussion}). That the scaleheight of the
thick radio disk is
larger than the effective beam proves the finding of extra-planar
emission and hence the existence of a radio halo. The scaleheights of
the thick disk at $\lambda 6.2\,\rm cm$, $\lambda 20\,\rm cm$, and
$\lambda 90\,\rm cm$ are smallest for the central strips where the
underlying SFR is highest. This important observation is the result of
synchrotron loss and is discussed in the next sections.
\begin{figure}[tbp]
\centering \resizebox{\hsize}{!}{ \includegraphics{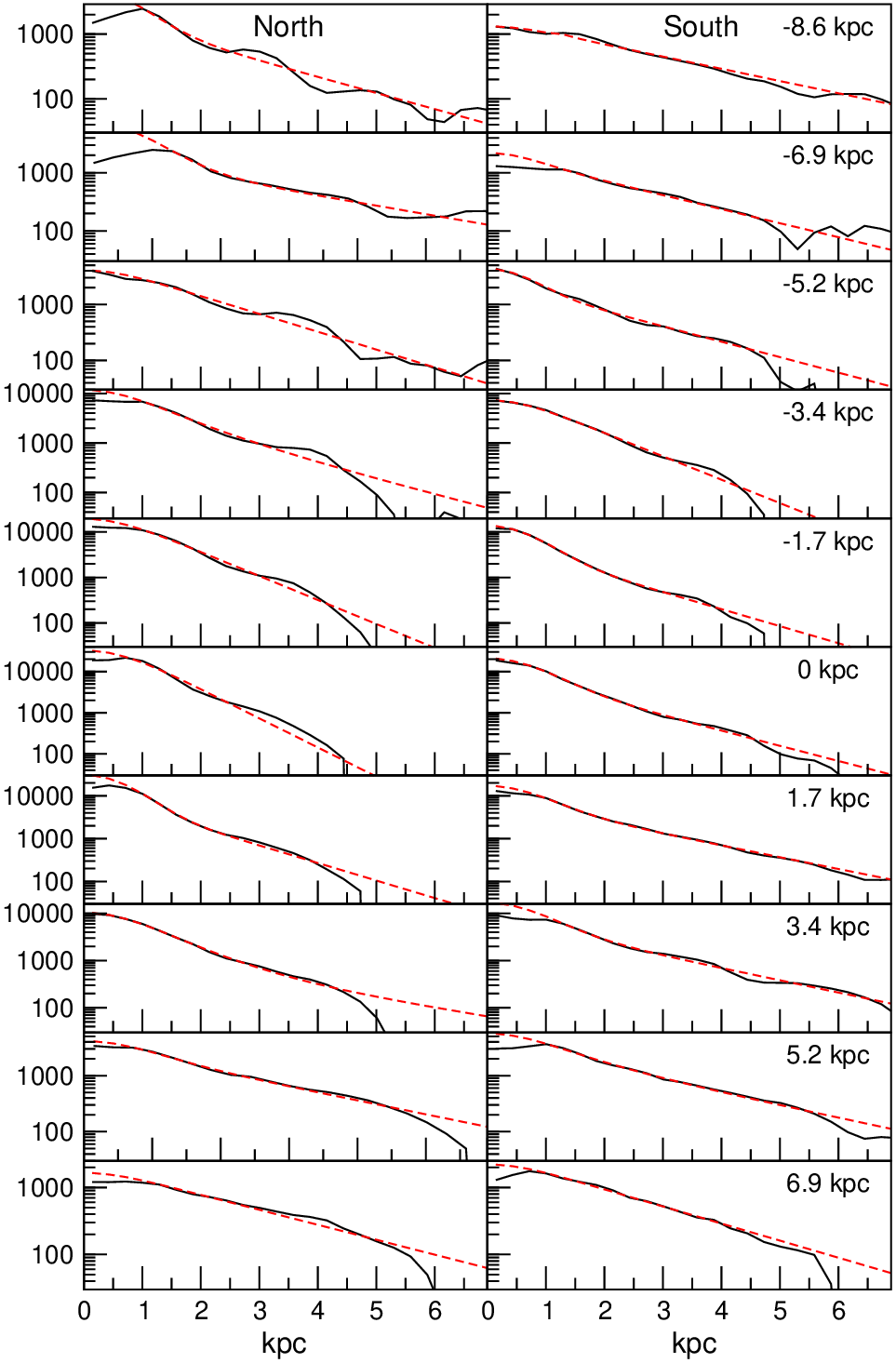}}
\caption[Exponential fits at $\lambda 6.2\,{\rm cm}$]{Total power
  emission profiles in $\mu\rm Jy/beam$ at $\lambda 6.2\,{\rm cm}$
  perpendicular to the disk. The solid line shows the measured profile
  and the dashed line shows the fitted profile consisting of two
  exponential functions convolved with the effective beam. The offset
  of the profiles along the major axis is denoted.}
\label{fig:i_x_6}
\end{figure}
\subsection{The total magnetic field strength}
\label{subsec:cr_magnetic_field}
\begin{figure}[tb]
\resizebox{\hsize}{!}{\includegraphics{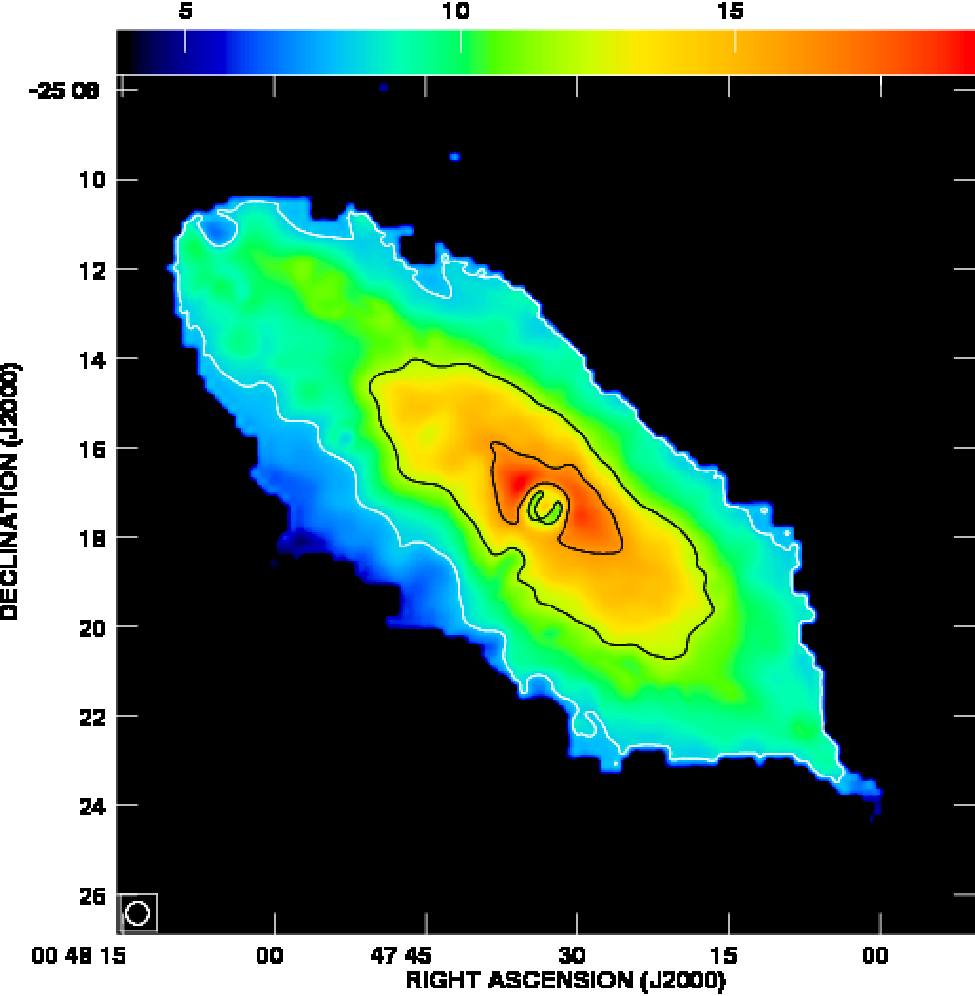}}
\caption[]{Map of the total magnetic
  field strength with a resolution of $30\arcsec$. Contours are at 8,
  12, 16, 20, and 24 $\times$ 
  $1\,{\rm\mu G}$.}
\label{fig:n253cm6ve_bf_b30}
\end{figure}

From the non-thermal total power emission one can determine the total
magnetic field strength assuming equipartition between the energy
density of the CRs and the magnetic field. Here we used the revised
equipartition formula by \citet{beck_05a} assuming a pathlength of
$6.5\,{\rm kpc}$, which is the full width to the half power of the
Gaussian fitted to the total power emission along the major
axis. Moreover, we assumed a non-thermal spectral index of
$\alpha_{\rm nt}=1.0$ (see Sect.~\ref{subsec:cr_si_total}), a
fractional polarization of 5.5\,\% (see Paper~II), and a number
density ratio $K_0=100$ of protons and electrons. We stress that the
last assumption might not be well fulfilled at large distances from
the disk, because the CREs suffer high synchrotron losses whereas the
CR protons are almost unaffected. Hence, the observed $K$ factor
increases with increasing distance from the CR sources which are
located in the disk. The magnetic field strength is then
underestimated by a small factor of $(K/K_0)^{1/4}$ \citep{beck_05a}.
The structure of the regular magnetic field in NGC\,253 is a
superposition of a disk and a vertical component (see Paper~II). The
disk component is a spiral with a constant pitch angle and the
inclination angle of the optical disk. The vertical component is
orientated perpendicular to the disk. The above presented parameters
as well as the geometry of the magnetic field were assumed to be
constant when computing the total magnetic field strength. To check
the sensitivity of this assumption we varied the parameters in a
certain range and observed its influence on the total magnetic field
strength. We used an integration path length between $5\,{\rm kpc}$
and $20\,{\rm kpc}$, a fractional polarization between 5\,\% and
20\,\%, a non-thermal spectral index between $0.8$ and $1.2$, and
assumed two structures for the magnetic field: disk-parallel or
vertical. The former one describes the geometry of the magnetic field
structure in the disk, where all possible magnetic field directions
along the line of sight are taken into account. The latter one
describes the geometry of the magnetic field in the halo, where a
constant magnetic field direction along the line of sight is
assumed. The variation of the total magnetic field strength in this
range of conditions was found to be $1.5\,\mu\rm G$, which we thus adopt as an
estimate for the systematic error of the total magnetic field
strength.
The total magnetic field strength was scaled with the non-thermal
synchrotron intensity $I_{\rm n}$ using the standard equipartition
formula \citep{beck_05a}:
\begin{equation}
B\propto I_{\rm n}^{1/(3+\alpha_{\rm nt})}.
\label{eq:cr_bl} 
\end{equation}
The distribution of the total magnetic field strength shown in
Fig.~\ref{fig:n253cm6ve_bf_b30} was calculated from the $\lambda
6.2\,\rm cm$ map. It ranges between $7\,{\rm \mu G}$ and $18\,{\rm\mu
  G}$ in the disk and is smaller in the halo. Note the subtracted
nucleus, which is highlighted by a notable residual hole in the center
of the map. The strip integration as described above applied to the
map of the total magnetic field strength was used in order to find the
maximum of the total magnetic field strength $B_0$ in each strip. This
gives a lower limit on the lifetime of the CREs as described in the
next section.
\citet{thompson_06a} argued that the magnetic field strength in
starburst galaxies may be underestimated using the equipartition
value. Our measured mean magnetic field strength of $12\,\mu \rm G$ in
the disk is accurately in equipartition with the mean gas
surface-density $\Sigma_{\rm g}=0.84\times10^{-2}\,\rm g\,cm^{-2}$
\citep{puche_91a}, so that the disk of NGC\,253 fits onto the
equipartition line, although its starburst nucleus does not (see Fig.~1
in \citet{thompson_06a}).
\subsection{Loss processes of the cosmic ray electrons}
\label{subsec:loss_processes_of_the_cosmic_ray_electrons}
Several processes are known through which the CREs lose energy, like
synchrotron radiation, inverse Compton (IC) losses, non-thermal
bremsstrahlung, ionization losses, and adiabatic losses. They all have
different timescales which determine the lifetimes of the CREs. In
order to calculate the timescales we adopt the equations by
\citet{pohl_90a}. The synchrotron lifetime of CREs is
\begin{equation}
 t_{\rm syn}=8.352\times 10^9\, E ({\rm GeV})^{-1} B ({\rm\mu
   G})^{-2}\,{\rm yr},
\label{eq:t_syn_gev}
\end{equation}
where $E$ is the CRE energy and $B$ is the total magnetic field
strength. In the following we use the relation between the CRE energy
$E$, the observation frequency $\nu$, and the total magnetic field
strength $B$ \citep{rybicki_86a}:
\begin{equation}
 E({\rm GeV})=(\nu/({\rm 16.1\,MHz}))^{1/2}B({\rm\mu G})^{-1/2}.
\label{eq:cr_cre_energy}
\end{equation}
For the $\lambda 6.2\,{\rm cm}$ observations with $\nu=4860\,{\rm
  MHz}$ and a typical magnetic field strength of $15\,{\rm\mu G}$ we
find $E=4.5\,{\rm GeV}$ and hence $t_{\rm syn}=8.2\times10^6\,{\rm
  yr}$. Inverse Compton (IC) radiation losses have the same energy
dependence as synchrotron losses and the corresponding lifetime is
\begin{equation}
 t_{\rm IC}=3.55\times10^8 (E\,({\rm GeV}))^{-1}(U_{\rm ph} / (10^{-12}\,{\rm
   erg\, cm^{-3}}))^{-1}\,{\rm yr},
\end{equation}
where $U_{\rm ph}$ is the total photon energy density. From the infrared
observations of \citet{radovich_01a} we find a total photon energy
density of $1.3\times10^{-12}\,{\rm erg\, cm^{-3}}$ which leads to
$t_{\rm IC}=6.1\times10^7\,{\rm yr}$. This is almost an order of magnitude
larger than the synchrotron lifetime.
For the non-thermal CRE bremsstrahlung lifetime we use
\begin{equation}
 t_{\rm brems}=3.96\times 10^{7}\, n({\rm cm^{-3}})^{-1}\,{\rm yr},
\end{equation}
where $n$ is the ISM gas density. The ionization losses lead to a CRE
lifetime of:
\begin{equation}
 t_{\rm ion}=9.5\times10^{6}\, n({\rm cm^{-3}})^{-1}E({\rm GeV})\,{\rm
   yr}.
\end{equation}
We now use a mean gas column density of $\Sigma_{\rm
  g}=4\,M_\odot\,\rm pc^{-2}$ for the total gas as determined by
\citet{puche_91a} from \ion{H}{I} observations. We define a cylinder
with the diameter of the disk $D_{25}=27.7\arcmin$ ($31.7\,{\rm kpc}$)
and adopt a height of $3.4\,{\rm kpc}$ which is twice the scaleheight
of the thick radio disk. The mean density of
$\rho=7.9\times10^{-26}\,{\rm g\, cm^{-3}}$ or $n\approx 0.05\,{\rm
  cm^{-3}}$ leads to a bremsstrahlung lifetime of $t_{\rm
  brems}=8\times10^8\,{\rm yr}$ and an ionization lifetime of $t_{\rm
  ion}=8.6\times10^8\,{\rm yr}$.
The adiabatic timescale
\begin{equation}
 t_{\rm ad}=3\left(\frac{d\rm v}{dz}\right)^{-1}.
\label{eq:adiabatic}
\end{equation}
depends only on the reciprocal velocity gradient. At this point we do not have
any information about the velocity profiles. We estimate, however,
the timescale of the adiabatic losses in
Sect.~\ref{subsec:cr_adiabatic_losses} as $3.6\times10^7\,\rm yr$, so
that the synchrotron losses and the adiabatic losses are the most
important loss processes of the CREs.
The local synchrotron lifetime calculated with Eq.~\ref{eq:t_syn_gev}
may be an overestimate, because
we observe at a certain location the superposition of locally produced
CRs and CRs, which were transported there from their origins. The
latter ones have already lost a fraction of their energy and have
shorter lifetimes than indicated by the local synchrotron losses. This
effect is large in the halo, where almost all CRs were transported
away from their origins. In the disk, where we observe the highest
fraction of locally produced CRs, this effect is small. Hence, we used
the minimum of the local synchrotron lifetime $t_{\rm syn,0}$
calculated from the maximum magnetic field strength $B_0$ in the disk
as the most accurate estimate.
Because $B$ decreases with $z$, $t_{\rm syn,0}$ is an
  underestimate. An improved estimate needs integration over the $B$
profile, which is, however, not known well enough. As a starting point
we assume an exponential vertical profile. As shown in
Appendix~\ref{app:t_syn} we find in this case that $t_{\rm syn,0}$ is
overestimated by no more than 10\,\% if we restrict the data to
$z<h_{\rm syn}$.
Summarizing this section, the CRE energy loss is dominated by the
synchrotron radiation and the adiabatic losses. Hence, we have to
combine the two energy losses in order to estimate the lifetime
$t_{\rm e}$ of the CREs, using:
\begin{equation}
\frac{1}{t_{\rm e}}=\frac{1}{t_{\rm syn,0}}+\frac{1}{t_{\rm ad}}.
\end{equation}
Since the adiabatic time scale is constant, the CRE lifetime is a
function of the synchrotron lifetime. This can be seen in
Fig.~\ref{fig:cr_scaleheights_90}, where we show the variation of
$B_0$ and $t_{\rm e}$ along the major axis. The magnetic field
strength has its maximum in the center of the disk where both the
synchrotron lifetime and the CRE lifetime have a minimum.
%
\subsection{The cosmic ray bulk speed}
\label{subsec:cr_bulk_speed}
%
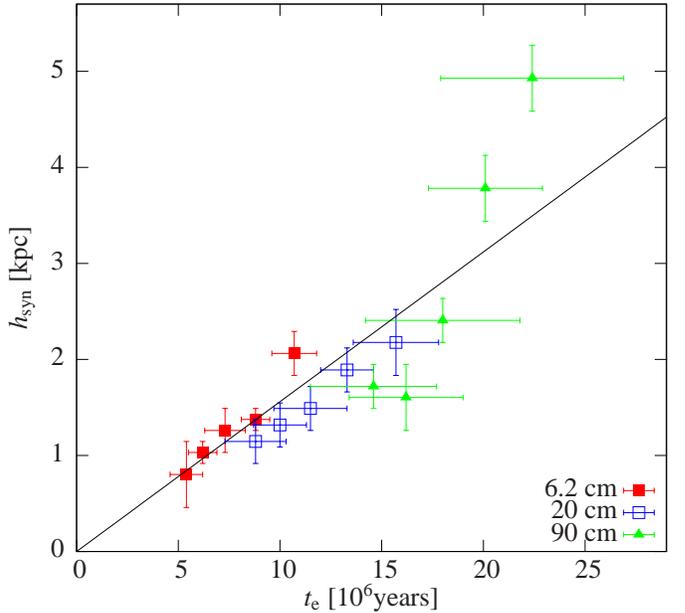
\begin{figure}[tb]
\resizebox{\hsize}{!}{
  \Large\input{0543fig14.tex}}
\caption{Scaleheight $h_{\rm syn}$ of the thick radio disk as a
  function of the CRE lifetime $t_{\rm e}$ in the northeastern halo
  (between $-6\,\rm kpc$ and $0\,\rm kpc$ offset along the major
  axis). The linear fit is the theoretical
  expectation for a convective CR transport with a constant CR bulk
  speed.}
\label{fig:h_sl_2e_E}
\end{figure}
\begin{figure}[tb]
\resizebox{\hsize}{!}{
  \Large\input{0543fig15.tex}}
\caption{Same as Fig.~\ref{fig:h_sl_2e_E} in the southwestern halo
  (between $1.5\,\rm kpc$ and $6\,\rm kpc$ offset along the major
  axis). The linear fit (dashed line) is
  the theoretical expectation for a convective CR transport with a
  constant CR bulk speed, while the square root fit (solid line) is for
  a diffusive CR transport (see text for details).}
\label{fig:h_sl_2e_W}
\end{figure}
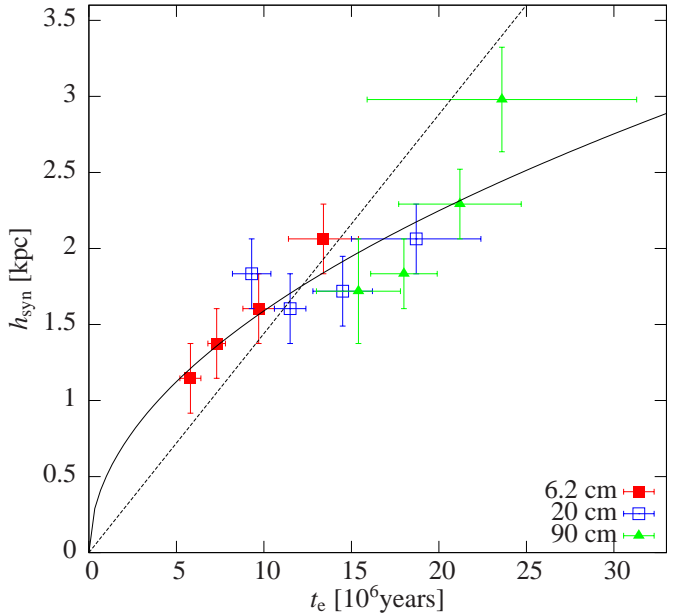
As described in Sect.~\ref{sec:cr_morphology} the total power emission
resembles a dumbbell shape at all observed wavelengths; the contour
lines show a strong thinning in the center close to the nucleus at the
brightest part of the disk. The analysis of the scaleheight of the
thick radio disk confirms this morphological analysis: it increases
with galactocentric radius with the minimum at the strip centered on
the nucleus. The opposite behavior is found for the magnetic field
strength, which decreases from the center to the outer part of the
disk. Hence, the scaleheight is small where the synchrotron lifetime
$t_{\rm syn,0}$ and hence the CRE lifetime $t_{\rm e}$ is small and
vice versa.
This relation motivates to calculate the average CR bulk speed,
\begin{equation}
{\rm v}=\frac{h_{\rm e}}{t_{\rm e}},
\label{eq:cr_bulk_speed}
\end{equation}
where $h_{\rm e}$ is the CRE scaleheight. Inserting the non-thermal synchrotron
intensity $I_{\rm n}\propto \exp(-z/h_{\rm syn})$ into
Eq.~\ref{eq:cr_bl} and using the equipartition condition $n_{\rm
  e}\propto B^2$ we find for the exponential profile of the CRE number density
\begin{equation}
n_{\rm e}\propto \left\{\exp\left(\frac{-z}{h_{\rm syn}}
\right)\right\}^\frac{\scriptstyle 2}{\rule{0in}{1ex}\scriptstyle 3+\alpha_{\rm nt}},
\end{equation}
where $h_{\rm syn}$ is the synchrotron scaleheight of the thick radio disk.
Thus, the CRE scaleheight is:
\begin{equation}
 h_{\rm e}=\frac{3+\alpha_{\rm nt}}{2}h_{\rm syn}.
\label{eq:cr_he}
\end{equation}
Since $\alpha_{\rm nt}=1$ the CRE scaleheight is twice the sclaleheight
of the thick radio disk.
In Fig.~\ref{fig:h_sl_2e_E} the scaleheight of the thick radio disk at
$\lambda 90\,\rm cm$, $\lambda 20\,\rm cm$, and $\lambda 6.2\,\rm cm$
is shown for each strip as a function of the CRE lifetime. We find a
(roughly) linear dependence between $h_{\rm syn}$ and $t_{\rm e}$. The
CR bulk speed in the northeastern halo is $300\pm30\,\rm
km\,s^{-1}$, where the error is dominated by the uncertainity in the
magnetic field strength and thus in the electron lifetime. The linear
least square fit has a reduced $\chi^2=3.6$. In the southwestern halo, we find a
flatter dependence between the scaleheight and the CRE lifetime
(Fig.~\ref{fig:h_sl_2e_W}) with a CR bulk speed of $250\pm30\,\rm
km\,s^{-1}$ with a ($\chi^2=2.7$). In this part of the halo the
scaleheight can be better fitted by square root function
($\chi^2=0.8$).
In the northeastern halo, the CR bulk speed is constant between
$\lambda90\,\rm cm$ and $\lambda6.2\,\rm cm$ and
does not vary with the CRE lifetime. This can be explained by a
convective CR transport, where the CRs and the magnetized ISM are
transported together. In this case, the CR bulk speed is independent
of the CR energy and hence in the northeastern halo the CR transport
is probably dominated by convection.
In the southwestern halo, the CR bulk speed decreases with increasing
CRE lifetime and increasing wavelength. This can be explained by a
diffusive CR transport where the CR bulk speed is ${\rm
  v}\approx\kappa/(\Gamma_{\rm c}\cdot h_{\rm e})$. Here, $\kappa$
is the diffusion coefficient and $\Gamma_{\rm c}=4/3$ is the adiabatic
index of the relativistic CR gas. Replacing the electron
scaleheight by $h_{\rm e}={\rm v}\cdot t_{\rm e}$ we obtain ${\rm
  v}\approx\sqrt{\kappa/(\Gamma_{\rm c}\cdot t_{\rm e})}$. Since the
diffusion coefficient is usually assumed to increase with the CR energy
($\kappa\propto E^{0.5}$), the CR bulk speed decreases both with increasing
wavelength and with increasing synchrotron lifetime. This agrees with
our observations and hence in the southwestern halo diffusion probably
plays a significant role. Using the square root fit of
Fig.~\ref{fig:h_sl_2e_W} we find a
diffusion coefficient of $\kappa=(2.0\pm0.2)\times10^{29}\,\rm
cm^{2}\,s^{-1}$ which is in the range of the values quoted by
\citet{breitschwerdt_93a}.
The relation between the scaleheight and the CRE lifetime explains the
observed pattern of the dumbbell total power emission. In the center
of the galaxy the magnetic field strength is highest and hence the
synchrotron lifetime is smallest; it hampers the transport of CREs
into the halo. This effect even overcomes the excess of emission in
the center. The contour lines thus show a steep gradient in the center
above and below the nucleus.
%
\subsection{Adiabatic losses}
\label{subsec:cr_adiabatic_losses}
\begin{figure}[t]
\centering \resizebox{\hsize}{!}{\Large
  \Large\input{0543fig16}}
\caption[]{Spectra of the diffuse emission in the northeastern halo,
  the central halo, and the southwestern halo.}
\label{fig:RS_flux}
\end{figure}
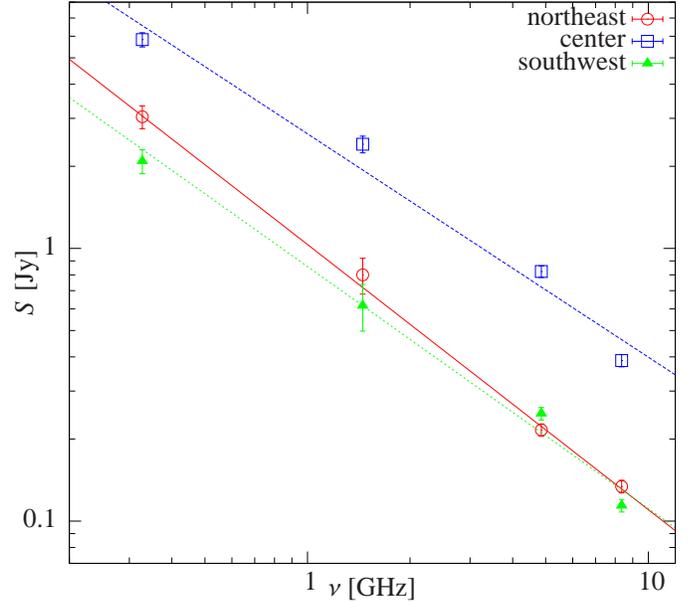

Using the information about the CR transport we can now calculate the
adiabatic loss timescale mentioned in
Sect.~\ref{subsec:loss_processes_of_the_cosmic_ray_electrons}. The CRs
are accelerated to their bulk speed of $300\,{\rm km\, s^{-1}}$ within
the CRE scaleheight of $3.6\,{\rm kpc}$. The adiabatic timescale is
thus $t_{\rm ad}=3.6\times10^7\,{\rm yr}$ using
Eq.~\ref{eq:adiabatic}, assuming a constant acceleration. This
timescale is similar to the synchrotron lifetime of the CREs. Hence,
the adiabatic losses have to be considered in particular for the
low-energy CREs emitting at $\lambda90\,{\rm cm}$. A transition of
adiabatic loss dominated CR transport to radiative dominated CR
transport should be visible in the spectrum as a steepening from a
flat spectral index at low frequencies to a steep spectral index at
high frequencies \citep{pohl_90c}. The break in the spectrum with a
steepening of $\Delta\alpha=0.5$ occurs at the frequency where the
adiabatic losses are equal to the radiation losses.
Since the CR transport is different for the northeastern and
southwestern halo it is worthwhile to investigate the spectra of the
extended flux density in these regions separately. For this we
integrated the flux density in three boxes with a width of
$5.2\,\rm kpc$ and a height of $5.7\,{\rm kpc}$ and an offset of
$-5.2\,\rm kpc$, $0\,\rm kpc$,  and $5.2\,\rm kpc$ along the major
axis. The corresponding spectra in Fig.~\ref{fig:RS_flux} steepen to
higher frequencies in particular the spectrum of the central
region. This indicates high adiabatic losses at $\lambda 90\,\rm cm$.
From linear fits we derive the spectral index as $0.97\pm0.05$,
$0.82\pm0.05$, and $0.89\pm0.05$ in the northeastern, central, and
southwestern halos, respectively.

\subsection{Cosmic ray spectral aging}
\label{subsec:cr_aging}
\begin{figure}[bthp]
\centering \resizebox{\hsize}{!}{ \includegraphics{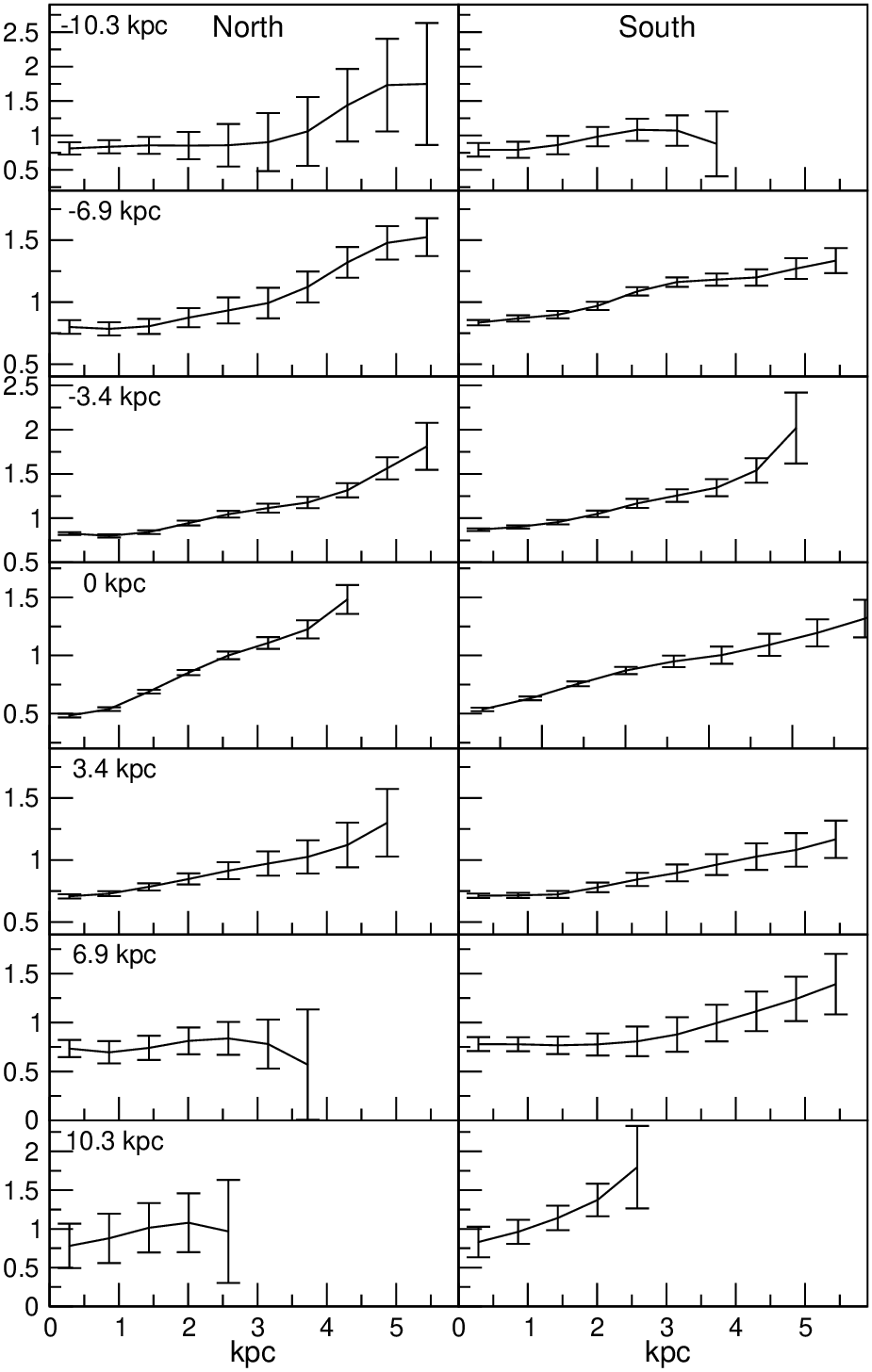}}
\caption[]{Profiles of the spectral index between $\lambda90\,\rm cm$
  and $\lambda 6.2\,\rm cm$ perpendicular to the disk as a function of
  distance from the major axis. The offset of the strips along the major
  axis is denoted.}
\label{fig:sp90_6_x}
\end{figure}
In the previous section we calculated the CR bulk speed from the
scaleheight of the CRE and their lifetime. Since the radiation power
of the synchrotron emission is proportional to $E^2$, the spectral
index steepens with time. This so-called CR {\it spectral aging} can
be used for an alternative approach to calculate the CR bulk speed as
we show in this section. 
A single CRE loses its energy via synchrotron radiation with a loss power of
\begin{equation}
\frac{dE}{dt} = -\frac{\sigma_{\rm
    T}cB_0^2}{8\pi}\left(\frac{E}{mc^2}\right)^2,
\label{eq:cr_losses2}
\end{equation}
where $\sigma_{\rm T}=6.65\times10^{-25}\,{\rm cm^2}$ is the Thomson
cross section, $B_0$ is the (constant) total magnetic field strength,
and $m=511\,{\rm keV}/c^2$ is the electron rest mass. The time
dependence of the CRE energy is $E(t)=E_0/(1+t/t_{\rm syn})$, where
the synchrotron lifetime $t_{\rm syn}$ is the time where a CRE has
lost half of its initial energy $E_0$.
The electron spectral index is defined by
\begin{equation}
\gamma=-\left.\log\frac{N_1}{N_2}\, \right / \, \log\frac{E_1}{E_2},
\label{eq:electron_spectral_index}
\end{equation}
where the number densities $N_1$ and $N_2$ are calculated for two
different CR energies $E_1$ and $E_2$. The electrons are
injected at $t=0$ and are assumed not to be reaccelerated in the halo. If
we neglect diffusion and assume a constant wind speed, the electrons
can be considered to reside in a box while they are losing their
energy due to synchrotron radiation. In this case their number
density can be expressed by \citep{lisenfeld_00a}
\begin{equation}
  N(E)=K_0 E^{-p}\left(1-\frac{t}{t_{\rm syn}}\right)^{p-1},
\end{equation}
where $p$ is the injection spectral index which is assumed to be about
$p\approx2$. Our model describes the geometry of a ``disk wind'' with a
vertical wind direction. Our analysis is similar to \citet{seaquist_91a} 
who made a spherical symmetric model for
the wind in the central part of M82. For the time derivative of the electron
spectral index we find for $t \to 0$
\begin{equation}
\frac{\Delta\gamma}{\Delta
  t}=(p-1)\left(\frac{1}{t_1}-\frac{1}{t_2}\right)\,\left /\,\log\frac{E_2}{E_1}\right.,
\end{equation}
where $t_1$ and $t_2$ denote synchrotron lifetimes. The
radio spectral index can be computed using $\alpha=(\gamma-1)/2$. With
Eqs.\ \ref{eq:t_syn_gev} and \ref{eq:cr_cre_energy} we can express the
time derivative of the radio spectral index by
\begin{eqnarray}
 \frac{\Delta\alpha}{\Delta t} & = & \frac{B_0({\mu\rm
     G})^{3/2}}{\ln(\nu_1/\nu_2)\cdot 8.352\times10^9\,{\rm
     yr}}\nonumber\\ & & \quad\times\left(\sqrt{\frac{\nu_2}{16.1\,{\rm
       MHz}}} - \sqrt{\frac{\nu_1}{16.1\,{\rm MHz}}}\right),
\end{eqnarray}
which only depends on the total magnetic field strength and the
observation frequencies $\nu_1$ and $\nu_2$. Thus, we get an
expression for the average CR bulk speed ${\rm v}=\Delta z/\Delta t$
\begin{equation}
{\rm v}=\left.\frac{\Delta\alpha}{\Delta
  t}\right / \frac{\Delta\alpha}{\Delta z},
\label{eq:cr_speed_a} 
\end{equation}
where ${\Delta\alpha}/{\Delta z}$ is the slope of the spectral index
profiles.
We applied a strip integration in order to calculate profiles of the
spectral index between $\lambda90\,\rm cm$ and $\lambda6.2\,\rm
cm$. The strip width was $180\arcsec$ and the pixel separation was
$15\arcsec$. The spectral index profiles are presented in
Fig.~\ref{fig:sp90_6_x} where the error bars were computed from
the noise in the strips, which is smaller than the noise in the map
(since the width of the strips is larger than the map resolution of
$70\arcsec$). The spectral index distribution shows an almost linear
steepening in some (but not all) of the strips. This can be explained
by a constant time derivative $\Delta\alpha/\Delta t$ and hence a
constant CR bulk speed $\rm v$.
The central strips at $0\,\rm kpc$ offset from the nucleus show
a very flat spectral index for $z<1\,\rm kpc$. This can be explained by a
thermal flattening in the nuclear region at $\lambda 90\,\rm cm$
\citep{carilli_96a}. We restricted the analysis of the spectral
index slope to heights larger than $1.4\,\rm kpc$ where we do not
expect any contribution of the thermal gas in the disk. The resulting CR
bulk speeds are shown in Table~\ref{tab:cr_alpha}, where the subscripts
``N'' and ``S'' denote the northern and and southern hemisphere,
respectively. The mean CR bulk speed derived from
the spectral index slope is $170\pm70\,{\rm km\, s^{-1}}$, to be
compared with the CR bulk speed obtained from the scaleheights of $\rm
v_{\lambda 6.2}=(300\pm 30)\,km\,s^{-1}$. Since the time derivative is
taken at $t=0$ this is a lower limit for the CR bulk speed.
\begin{table}[htb]
\caption[]{Average CR bulk speeds obtained from the radio spectral
  index aging.}
\begin{center}
\begin{tabular}{llllll}
\hline\hline $x$ & $\left.\frac{\textstyle \Delta\alpha}{\textstyle
  \Delta z}\right|_{\rm N}$ \T\B&
$\left.\frac{\textstyle \Delta\alpha}{\textstyle\Delta z}\right|_{\rm
  S}$ & $B_0$ & ${\rm v}_{\rm N}$ & ${\rm
  v}_{\rm S}$\\
$[\rm kpc]$ \B\T & $[\rm kpc^{-1}]$ & $[\rm kpc^{-1}]$ & $[\rm\mu G]$
& $[\rm km\,s^{-1}]$ & $[\rm km\,s^{-1}]$\\\hline
$-6.9$ \T & -0.22 & -0.12 & 11 & 107 & 196\\ 
$-3.4$ &-0.26 & -0.31 & 14 & 130 & 109\\ 
$0$ & -0.30 & -0.26 & 17 & 150 & 174\\ 
$3.4$ & -0.16 & -0.12 & 15 & 234 & 312\\ 
$6.9$ \B & -0.10 & -0.20 & 10 & 204 & 102\\\hline
\end{tabular}
\end{center}
\label{tab:cr_alpha}
\end{table}
\section{Discussion}
\label{sec:cr_discussion}
The CR distribution to constrain the transport of CRs
from the disk into the halo. The local CRE lifetime is
dominated by synchrotron losses and depends on the magnetic field strength
in the disk which is highest in the central part. The dependence of
the scaleheight on the CRE lifetime, as evident from the dumbbell shaped halo, 
requires a vertical CR transport.
The CR bulk speed, defined by the ratio of the CRE scaleheight to
their lifetime, is constant in the northeastern halo with
$300\pm30\,\rm km\,s^{-1}$ and remarkably similar when applied to our
observations at three different wavelengths. Thus, the local CR
  bulk speed is independent of the CR energy, the star-formation
  rate (SFR), and the magnetic field strength. Our second method to
determine the CR bulk speed is to use the spectral aging of the
CREs. The steepening of the observed radio spectral index with
increasing distance from the galactic midplane results in a lower
limit for the CR bulk speed with $170\pm70\,\rm
km\,s^{-1}$. \cite{zirakashvili_06a} made an analytical model for the
CR transport in the nuclear outflow.  They found by comparison with
radio continuum data a velocity of $300\,\rm km\,s^{-1}$ near the disk
and $900\,\rm km\,s^{-1}$ as terminal velocity.
\citet{dahlem_95a} and \citet{dahlem_06a} studied the CR distribution
in edge-on galaxies. They found a tight and linear correlation between
the radial extent of radio halos and the level of star-formation
activity in the underlying disk. However, radio halos can be better
studied by their vertical intensity distributions. Whereas the
amplitude of the vertical profile depends on the overall radio
intensity of the disk, related to the SFR, its
scaleheight is determined by the outflow velocity and the CRE
lifetime. The observed vertical extent also depends on the
signal-to-noise ratio of the measurements. Only the scaleheight is
suited to characterize the formation of radio halos.
Global radio scaleheights observed at $\lambda 6\,\rm cm$ in three
edge-on galaxies with very different global SFRs were found to be
similar, about $1.8\,\rm kpc$ for the thick radio disk
\citep{dumke_98a, krause_04a}. This agrees well with our result of
$1.7\,\rm kpc$ of NGC\,253 which has the highest SFR of this
sample. Our Eq.~\ref{eq:cr_bulk_speed} indicates that a constant CRE
scaleheight $h_{\rm e}$ in galaxies with different SFRs and hence
different CRE lifetimes $t_{\rm e}$ can only be achieved if the global
CR bulk speed $\rm v$ increases with global SFR. As all four galaxies
show a dumbbell shaped radio halo, the local scaleheight depends on
the local magnetic field strength.
Since we postulate an outflow, we analyse the pressure contributions
of the different ISM phases. These are the thermal gas, the kinetic
energy, and the magnetic field. The thermal pressure of the halo
gas is $P_{\rm therm}=2n_{\rm e}k_{\rm B}T=7\times10^{-12}\,{\rm erg\,
  cm^{-3}}$, where we used an electron density of $n_{\rm
  e}=8\times10^{-3}\,{\rm cm^{-3}}$ and a temperature of
$T=3\times10^6\,{\rm K}$ \citep{strickland_02a}. The ram pressure of
the outflow is $P_{\rm ram}=1/2n_{\rm e}m_{\rm p}{\rm
  v}^2=6\times10^{-12}\,{\rm erg\, cm^{-3}}$, where $m_{\rm p}$ is the
proton mass and ${\rm v}=300\,{\rm km\, s^{-1}}$ the wind
velocity. Assuming a typical magnetic field strength of $B=15\,\mu\rm
G$, we find a magnetic pressure of $P_{\rm
  mag}=B^2/(8\pi)=9\times10^{-12}\,{\rm erg\, cm^{-3}}$.
If the gas in the disk were in a hydrostatic equilibrium, the gas
pressure would be $P_{\rm grav}\approx\pi G\Sigma_{\rm g}\Sigma_{\rm
  t}$. We obtain $P_{\rm grav}=1.3\times10^{-11}\,\rm erg\, cm^{-3}$,
where we inserted $\Sigma_{\rm g}=4\,\rm M_\odot\, pc^{-2}$ for
the gas surface-density \citep{puche_91a}, and $\Sigma_{\rm
  m}=7.44\times10^{-2}\,\rm g\,cm^{-2}$ for the total mass
surface-density \citep{pence_81a}. We note, that the thermal pressure,
the ram pressure, and the magnetic pressure are of the same order of
magnitude. Since the sum of the thermal pressure, the ram pressure,
and the magnetic pressure is $2.2\times10^{-11}\,\rm erg\,cm^{-3}$,
this may indicate that the disk is not in hydrostatic equilibrium and
allows outflow.
If the outflow traced by the CR transport is a galactic wind, it
should finally leave the gravitational potential of the galaxy. In
this case the CR bulk speed must exceed the escape velocity of ${\rm
  v}_{\rm esc}=2^{1/2}\cdot \Omega r$ \citep{veilleux_05a}, where
$\Omega r$ is the rotation velocity of the disk. Of course, the
outflow may accelerate further away from the disk, below our
sensitivity limit, and thus a galactic wind is possible even if the
observed CR bulk speed is too low. On the other hand, a CR bulk speed
above the escape velocity supports the galactic wind scenario, because
it will certainly escape from the gravitational potential, unless the
outflow decelerates in the halo. Using the rotation speed of
$200\,{\rm km\,s^{-1}}$ measured by \citet{pence_81a}, we find an
escape velocity of $280\,{\rm km\,s^{-1}}$, which agrees within the
errors with our measured CR bulk speed. Hence, a galactic wind is in
NGC\,253 likely and the CR transport shows the presence of a ``disk
wind'' in the northeastern halo.
The disk wind can efficiently transport gas into the halo. The
transport should be more efficient in the convective northeastern
halo. Indeed, there is significantly
more neutral hydrogen gas in the northeastern halo, as \ion{H}{I}
observations by \citet{boomsma_05a} show. A similar distribution
is found for the ionized hydrogen, where prominent H$\alpha$
plumes are found only in the northeastern halo
\citep{hoopes_96a}. This picture is corroborated by huge lobes of soft
X-ray emission in the northeastern halo, which are absent in the
southwestern halo \citep{pietsch_00a}. The discrepancy between the
northeastern and the southwestern halo may be explained by the level
of star-formation activity in the disk. Both the H$\alpha$ emission
and radio continuum emission is stronger in the northeastern part of
the disk. This hints to a higher SFR that can more easily drive the
disk wind against the gravitational pull.
The disk wind may finally escape from the gravitational potential to
the intergalactic medium (IGM). From the observations we cannot make a
definite conclusion because we do not observe the transition between
the radio halo and the IGM further away from the detected radio
halo. With our observations we are not able to conclude whether the disk wind
accelerates or decelerates within distance from the disk. With low
frequency observations the scaleheight of the synchrotron emission is
expected to be larger and hence the radio halo could be observed to
larger vertical heights. Such data could allow us to determine
whether the disk wind accelerates as expected for a CR driven
wind. Although LOFAR observations would be ideal for this purpose, the
low declination of NGC\,253 ($-25^\circ$) requires to use the EVLA or
the future SKA at frequencies as low as possible.
\section{Conclusions}
\label{sec:cr_conclusions}
Our radio continuum study of NGC\,253 shows an extended radio halo at
the wavelengths from $\lambda 90\,{\rm cm}$ to $\lambda 3.6\,{\rm
  cm}$. Because studies of NGC\,253 need a high dynamic range, it is a
difficult target for radio continuum observations. We treated the high
dynamic range with a specially tailored data reduction technique. Our
maps presented here are thus essentially noise limited. The
combination of our new VLA mosaic map with the single-dish Effelsberg
data in order to fill in the missing zero-spacing flux density allowed
us to study the extra-planar continuum emission with high resolution
and sensitivity. They are used to put detailed constraints on the CR
transport, which lead to our main conclusions:
\begin{enumerate}
\item The scaleheight of the thick radio disk in the northeastern halo
  is a linear function of the CRE lifetime in the underlying disk at
  $\lambda90\,\rm cm$, $\lambda 20\,\rm cm$, and $\lambda 6.2\,\rm cm$. This
  can be understood if the CR transport proceeds preferentially in
  the vertical direction.
\item The steepening of the spectral index, which is a linear function
  of $z$, can be explained by strong synchrotron losses of the
  CREs. The contribution of ionization losses and bremsstrahlung
  losses can be neglected. The adiabatic losses are of the same order
  of magnitude as the synchrotron losses at $\lambda 20\,\rm cm$ but dominate at
  $\lambda 90\,\rm cm$.
\item We computed the CR bulk speed as the ratio of the total power
  scaleheight and the CRE lifetime. The CR bulk speed in the
  northeastern halo of
  $300\pm 30\,{\rm km\, s^{-1}}$ is similar to the escape velocity
  of $280\,{\rm km\,s^{-1}}$, indicating the
  existence of a disk wind which escapes from the gravitational
  potential. The local CR bulk speed is independent of the CR energy,
  the SFR, and the magnetic field strength.
\item The CR transport is mainly convective in the northeastern halo
  while in the southwestern halo the diffusion of CRs is
  significant. We measure an average diffusion coefficient
  of $2.0\pm0.3\times10^{29}\,\rm cm^2\,s^{-1}$
\item The disk wind is the driving source for the luminous gas in the halo. In
  the convective northeastern halo the gas is transported more
  efficiently from the disk into the halo than in the diffusive
  southwestern halo. This can explain the different amounts of
  extra-planar gas in these two halo parts as observed in
  \ion{H}{I}, H$\alpha$, and soft X-rays.
\end{enumerate}
The CR transport is always the superposition of convection and
diffusion. The data presented here provide a basis for more detailed
calculations using the combined diffusion and convection
equation. These calculations should include vertical profiles of the
magnetic field strength, the CR number density, and the velocity of the
galactic wind.
\begin{acknowledgements}
We appreciate the comments of our referee, Timothy Paglione, which
improved the paper significantly. We thank Michael Dumke and the
Effelsberg staff for help with the observations. It is our pleasure to
thank Reinhard Schlickeiser and Dieter Breitschwerdt for inspiring
discussions. We are grateful to Enno Middelberg and Wolfgang Reich for
valuable comments on the manuscript.\\ VH would like to thank the
organizers of the Graduiertenkolleg GRK 787 and the
Sonderforschungsbereich SFB 591 for support and funding during the
course of his PhD.  The GRK 787 ``Galaxy groups as laboratories for
baryonic and dark matter'' and the SFB 591 ``Universal properties of
non-equilibrium plasmas`` were funded by the Deutsche
Forschungsgemeinschaft (DFG).
\end{acknowledgements}
\appendix
\section{The synchrotron lifetime within an exponential magnetic field profile}
\label{app:t_syn}
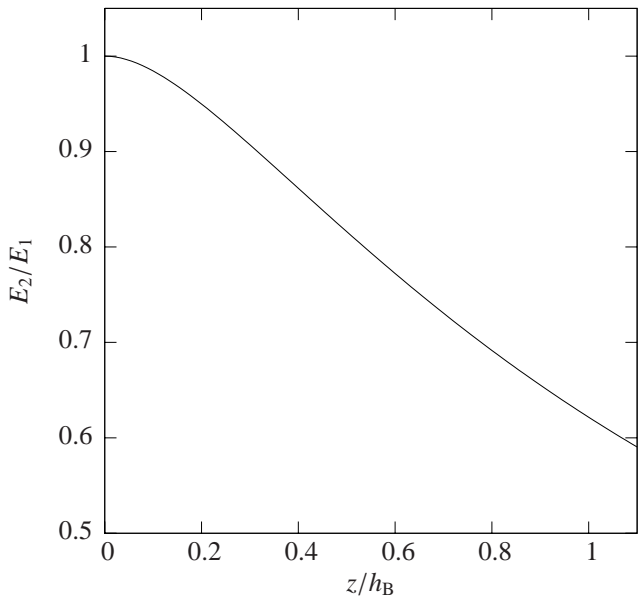
\begin{figure}[htbp]
\resizebox{\hsize}{!}{
  \Large\input{0543figA1}}
\caption[]{Ratio of the two energies $E_{\rm B0}/E_{\rm B}$ as a
  function of $z$. For $E_{\rm B}$ the magnetic field has an exponential
  distribution with a scaleheight $h_{\rm B}$ while for $E_{\rm B0}$ it is
  constant.}
\label{fig:t_syn}
\end{figure}

In order to calculate the CR bulk speed in
Sect.~\ref{subsec:cr_bulk_speed} we used the synchrotron
lifetime of the electrons. Since we do not know the vertical
distribution of the magnetic field, we assumed a constant
magnetic field strength. In reality, however, we expect the magnetic
field strength to have a vertical, decreasing profile. In this section
we investigate the difference between the synchrotron lifetime for
a constant magnetic field and for an exponential vertical profile.
The exponential
profile of the magnetic field reads
\begin{equation}
 B(z)=B_0 \exp(-z/h_{\rm B}),
\end{equation}
where $B_0$ is the maximum magnetic field strength and $h_{\rm B}$ the
scaleheight of the magnetic field. Replacing $t=z/\rm v$, where we
assume a constant CR bulk speed, the differential of the energy is:
\begin{equation}
 dE= -\frac{\sigma_{\rm
     T}cB(z)^2}{8\pi}\left(\frac{E(z)}{mc^2}\right)^2 \frac{dz}{\rm
   v}.
\end{equation}
We integrate this differential equation by separating the
variables with:
\begin{equation}
 \frac{dE}{E^2}=-\frac{\sigma_{\rm T}B(z)^2}{8\pi m^2c^3\rm v}dz.
\end{equation}
Now we integrate from $z=0$ to $z$:
\begin{equation}
 \int_{E_0}^{E}\frac{dE}{E^2}=\int_{z=0}^{z}-\frac{\sigma_{\rm
     T}B_0^2}{8\pi m^2c^3\rm v}\exp(-2z/h_{\rm B})dz,
\end{equation}
where we inserted the exponential profile of $B$. Finally,
we find for the vertical CRE energy profile
\begin{equation}
 E(z)=\frac{E_0}{\displaystyle 1 + \frac{E_0h_{\rm B}\sigma_{\rm
       T}B_0^2}{\displaystyle 16\pi m^2c^3\rm v}\cdot\left(1-\exp(-2z/h_{\rm
     B})\right)}.
\end{equation}
If we use the definition for the synchrotron lifetime for a constant
field strength $B_0$
\begin{equation}
 t_{\rm syn,0}=\frac{8\pi m^2c^3\null}{E_0B_0^2}
\end{equation}
we find
\begin{equation}
 E_{\rm B}(z)=\frac{E_0}{1+\frac{\displaystyle h_{\rm B}}{\displaystyle 2
     t_{\rm syn,0}\rm{v}}\cdot\left(1-\exp(-2z/h_{\rm B})\right)}
\end{equation}
For the case of a constant magnetic field strength we have as
energy profile:
\begin{equation}
 E_{\rm B0}(z)=\frac{E_0}{1+\frac{\displaystyle z}{\displaystyle
     t_{\rm syn,0}\rm {v}}}.
\end{equation}
Clearly, the larger the scaleheight of the magnetic field, the smaller
is the difference between the two CRE energy profiles. We now use
$h_{\rm B}=4\cdot h_{\rm syn}$ as a lower limit due to the
equipartition condition (see Eq.~\ref{eq:cr_bl}), where we set
$\alpha_{\rm nt}=1$. In Fig.~\ref{fig:t_syn} we plot the ratio of the
energies $E_{\rm B0}/E_{\rm B}$ as a function of $z$ in units of $h_{\rm B}$. Any
estimate of the error made using a constant magnetic field instead of
a exponential magnetic field must be averaged using the weight of the
points. If we restrict $z<h_{\rm syn}$ and take into account that the
inner points have larger weights (due to the higher synchrotron
emission), the error estimate is around 10\,\%. Hence, in our
case the assumption of a constant magnetic field strength for the
calculation of the CRE energy and hence of the synchrotron
lifetime in
Sect.~\ref{subsec:loss_processes_of_the_cosmic_ray_electrons} is acceptable. 
\end{document}

%% file: 0543fig11.tex
\begingroup
  \makeatletter
  \providecommand\color[2][]{%
    \GenericError{(gnuplot) \space\space\space\@spaces}{%
      Package color not loaded in conjunction with
      terminal option `colourtext'%
    }{See the gnuplot documentation for explanation.%
    }{Either use 'blacktext' in gnuplot or load the package
      color.sty in LaTeX.}%
    \renewcommand\color[2][]{}%
  }%
  \providecommand\includegraphics[2][]{%
    \GenericError{(gnuplot) \space\space\space\@spaces}{%
      Package graphicx or graphics not loaded%
    }{See the gnuplot documentation for explanation.%
    }{The gnuplot epslatex terminal needs graphicx.sty or graphics.sty.}%
    \renewcommand\includegraphics[2][]{}%
  }%
  \providecommand\rotatebox[2]{#2}%
  \@ifundefined{ifGPcolor}{%
    \newif\ifGPcolor
    \GPcolorfalse
  }{}%
  \@ifundefined{ifGPblacktext}{%
    \newif\ifGPblacktext
    \GPblacktexttrue
  }{}%
  \let\gplgaddtomacro\g@addto@macro
  \gdef\gplbacktext{}%
  \gdef\gplfronttext{}%
  \makeatother
  \ifGPblacktext
    \def\colorrgb#1{}%
    \def\colorgray#1{}%
  \else
    \ifGPcolor
      \def\colorrgb#1{\color[rgb]{#1}}%
      \def\colorgray#1{\color[gray]{#1}}%
      \expandafter\def\csname LTw\endcsname{\color{white}}%
      \expandafter\def\csname LTb\endcsname{\color{black}}%
      \expandafter\def\csname LTa\endcsname{\color{black}}%
      \expandafter\def\csname LT0\endcsname{\color[rgb]{1,0,0}}%
      \expandafter\def\csname LT1\endcsname{\color[rgb]{0,1,0}}%
      \expandafter\def\csname LT2\endcsname{\color[rgb]{0,0,1}}%
      \expandafter\def\csname LT3\endcsname{\color[rgb]{1,0,1}}%
      \expandafter\def\csname LT4\endcsname{\color[rgb]{0,1,1}}%
      \expandafter\def\csname LT5\endcsname{\color[rgb]{1,1,0}}%
      \expandafter\def\csname LT6\endcsname{\color[rgb]{0,0,0}}%
      \expandafter\def\csname LT7\endcsname{\color[rgb]{1,0.3,0}}%
      \expandafter\def\csname LT8\endcsname{\color[rgb]{0.5,0.5,0.5}}%
    \else
      \def\colorrgb#1{\color{black}}%
      \def\colorgray#1{\color[gray]{#1}}%
      \expandafter\def\csname LTw\endcsname{\color{white}}%
      \expandafter\def\csname LTb\endcsname{\color{black}}%
      \expandafter\def\csname LTa\endcsname{\color{black}}%
      \expandafter\def\csname LT0\endcsname{\color{black}}%
      \expandafter\def\csname LT1\endcsname{\color{black}}%
      \expandafter\def\csname LT2\endcsname{\color{black}}%
      \expandafter\def\csname LT3\endcsname{\color{black}}%
      \expandafter\def\csname LT4\endcsname{\color{black}}%
      \expandafter\def\csname LT5\endcsname{\color{black}}%
      \expandafter\def\csname LT6\endcsname{\color{black}}%
      \expandafter\def\csname LT7\endcsname{\color{black}}%
      \expandafter\def\csname LT8\endcsname{\color{black}}%
    \fi
  \fi
  \setlength{\unitlength}{0.0500bp}%
  \begin{picture}(14400.00,10080.00)%
    \gplgaddtomacro\gplbacktext{%
      \csname LTb\endcsname%
      \put(822,5852){\makebox(0,0)[r]{\strut{} 0}}%
      \put(822,6362){\makebox(0,0)[r]{\strut{} 1}}%
      \put(822,6873){\makebox(0,0)[r]{\strut{} 2}}%
      \put(822,7383){\makebox(0,0)[r]{\strut{} 3}}%
      \put(822,7894){\makebox(0,0)[r]{\strut{} 4}}%
      \put(822,8404){\makebox(0,0)[r]{\strut{} 5}}%
      \put(822,8915){\makebox(0,0)[r]{\strut{} 6}}%
      \put(822,9425){\makebox(0,0)[r]{\strut{} 7}}%
      \put(822,9936){\makebox(0,0)[r]{\strut{} 8}}%
      \put(1291,5600){\makebox(0,0){\strut{}-8}}%
      \put(2062,5600){\makebox(0,0){\strut{}-6}}%
      \put(2832,5600){\makebox(0,0){\strut{}-4}}%
      \put(3603,5600){\makebox(0,0){\strut{}-2}}%
      \put(4373,5600){\makebox(0,0){\strut{} 0}}%
      \put(5144,5600){\makebox(0,0){\strut{} 2}}%
      \put(5914,5600){\makebox(0,0){\strut{} 4}}%
      \put(6685,5600){\makebox(0,0){\strut{} 6}}%
      \put(7154,5852){\makebox(0,0)[l]{\strut{} 0}}%
      \put(7154,7048){\makebox(0,0)[l]{\strut{} 10}}%
      \put(7154,8244){\makebox(0,0)[l]{\strut{} 20}}%
      \put(7154,9440){\makebox(0,0)[l]{\strut{} 30}}%
      \put(500,7945){\rotatebox{90}{\makebox(0,0){$h_{\rm syn}~[{\rm kpc}]$}}}%
      \put(3988,5362){\makebox(0,0){$x~[{\rm kpc}]$}}%
      \put(1636,9688){\makebox(0,0){\strut{}\framebox{$\lambda 90\,\rm cm$}}}%
    }%
    \gplgaddtomacro\gplfronttext{%
      \csname LTb\endcsname%
      \put(5347,9779){\makebox(0,0)[r]{\strut{}$h_{\rm syn}$}}%
      \csname LTb\endcsname%
      \put(5347,9464){\makebox(0,0)[r]{\strut{}$B_0$}}%
      \csname LTb\endcsname%
      \put(5347,9149){\makebox(0,0)[r]{\strut{}$t_{\rm syn}$}}%
    }%
    \gplgaddtomacro\gplbacktext{%
      \csname LTb\endcsname%
      \put(822,812){\makebox(0,0)[r]{\strut{} 0}}%
      \put(822,1487){\makebox(0,0)[r]{\strut{} 1}}%
      \put(822,2162){\makebox(0,0)[r]{\strut{} 2}}%
      \put(822,2837){\makebox(0,0)[r]{\strut{} 3}}%
      \put(822,3513){\makebox(0,0)[r]{\strut{} 4}}%
      \put(822,4188){\makebox(0,0)[r]{\strut{} 5}}%
      \put(822,4863){\makebox(0,0)[r]{\strut{} 6}}%
      \put(906,560){\makebox(0,0){\strut{}-10}}%
      \put(1555,560){\makebox(0,0){\strut{}-8}}%
      \put(2204,560){\makebox(0,0){\strut{}-6}}%
      \put(2853,560){\makebox(0,0){\strut{}-4}}%
      \put(3501,560){\makebox(0,0){\strut{}-2}}%
      \put(4150,560){\makebox(0,0){\strut{} 0}}%
      \put(4799,560){\makebox(0,0){\strut{} 2}}%
      \put(5448,560){\makebox(0,0){\strut{} 4}}%
      \put(6097,560){\makebox(0,0){\strut{} 6}}%
      \put(6746,560){\makebox(0,0){\strut{} 8}}%
      \put(7154,812){\makebox(0,0)[l]{\strut{} 0}}%
      \put(7154,1763){\makebox(0,0)[l]{\strut{} 5}}%
      \put(7154,2715){\makebox(0,0)[l]{\strut{} 10}}%
      \put(7154,3666){\makebox(0,0)[l]{\strut{} 15}}%
      \put(7154,4617){\makebox(0,0)[l]{\strut{} 20}}%
      \put(500,2905){\rotatebox{90}{\makebox(0,0){$h_{\rm syn}~[{\rm kpc}]$}}}%
      \put(3988,322){\makebox(0,0){$x~[{\rm kpc}]$}}%
      \put(1636,4648){\makebox(0,0){\strut{}\framebox{$\lambda 6.2\,\rm cm$}}}%
    }%
    \gplgaddtomacro\gplfronttext{%
    }%
    \gplgaddtomacro\gplbacktext{%
      \csname LTb\endcsname%
      \put(8022,5852){\makebox(0,0)[r]{\strut{} 0}}%
      \put(8022,6362){\makebox(0,0)[r]{\strut{} 1}}%
      \put(8022,6873){\makebox(0,0)[r]{\strut{} 2}}%
      \put(8022,7383){\makebox(0,0)[r]{\strut{} 3}}%
      \put(8022,7894){\makebox(0,0)[r]{\strut{} 4}}%
      \put(8022,8404){\makebox(0,0)[r]{\strut{} 5}}%
      \put(8022,8915){\makebox(0,0)[r]{\strut{} 6}}%
      \put(8022,9425){\makebox(0,0)[r]{\strut{} 7}}%
      \put(8022,9936){\makebox(0,0)[r]{\strut{} 8}}%
      \put(8448,5600){\makebox(0,0){\strut{}-8}}%
      \put(9133,5600){\makebox(0,0){\strut{}-6}}%
      \put(9818,5600){\makebox(0,0){\strut{}-4}}%
      \put(10503,5600){\makebox(0,0){\strut{}-2}}%
      \put(11188,5600){\makebox(0,0){\strut{} 0}}%
      \put(11873,5600){\makebox(0,0){\strut{} 2}}%
      \put(12558,5600){\makebox(0,0){\strut{} 4}}%
      \put(13243,5600){\makebox(0,0){\strut{} 6}}%
      \put(13928,5600){\makebox(0,0){\strut{} 8}}%
      \put(14354,5852){\makebox(0,0)[l]{\strut{} 0}}%
      \put(14354,6600){\makebox(0,0)[l]{\strut{} 5}}%
      \put(14354,7347){\makebox(0,0)[l]{\strut{} 10}}%
      \put(14354,8095){\makebox(0,0)[l]{\strut{} 15}}%
      \put(14354,8842){\makebox(0,0)[l]{\strut{} 20}}%
      \put(14354,9590){\makebox(0,0)[l]{\strut{} 25}}%
      \put(7700,7945){\rotatebox{90}{\makebox(0,0){$h_{\rm syn}~[{\rm kpc}]$}}}%
      \put(11188,5362){\makebox(0,0){$x~[{\rm kpc}]$}}%
      \put(8836,9688){\makebox(0,0){\strut{}\framebox{$\lambda 20\,\rm cm$}}}%
    }%
    \gplgaddtomacro\gplfronttext{%
    }%
    \gplgaddtomacro\gplbacktext{%
      \csname LTb\endcsname%
      \put(8190,812){\makebox(0,0)[r]{\strut{} 0}}%
      \put(8190,1763){\makebox(0,0)[r]{\strut{} 0.5}}%
      \put(8190,2715){\makebox(0,0)[r]{\strut{} 1}}%
      \put(8190,3666){\makebox(0,0)[r]{\strut{} 1.5}}%
      \put(8190,4617){\makebox(0,0)[r]{\strut{} 2}}%
      \put(8607,560){\makebox(0,0){\strut{}-8}}%
      \put(9273,560){\makebox(0,0){\strut{}-6}}%
      \put(9940,560){\makebox(0,0){\strut{}-4}}%
      \put(10606,560){\makebox(0,0){\strut{}-2}}%
      \put(11272,560){\makebox(0,0){\strut{} 0}}%
      \put(11938,560){\makebox(0,0){\strut{} 2}}%
      \put(12604,560){\makebox(0,0){\strut{} 4}}%
      \put(13271,560){\makebox(0,0){\strut{} 6}}%
      \put(13937,560){\makebox(0,0){\strut{} 8}}%
      \put(14354,812){\makebox(0,0)[l]{\strut{} 0}}%
      \put(14354,1763){\makebox(0,0)[l]{\strut{} 5}}%
      \put(14354,2715){\makebox(0,0)[l]{\strut{} 10}}%
      \put(14354,3666){\makebox(0,0)[l]{\strut{} 15}}%
      \put(14354,4617){\makebox(0,0)[l]{\strut{} 20}}%
      \put(7700,2905){\rotatebox{90}{\makebox(0,0){$h_{\rm syn}~[{\rm kpc}]$}}}%
      \put(11272,322){\makebox(0,0){$x~[{\rm kpc}]$}}%
      \put(8920,4648){\makebox(0,0){\strut{}\framebox{$\lambda 3.6\,\rm cm$}}}%
    }%
    \gplgaddtomacro\gplfronttext{%
    }%
    \gplbacktext
    \put(0,0){\includegraphics{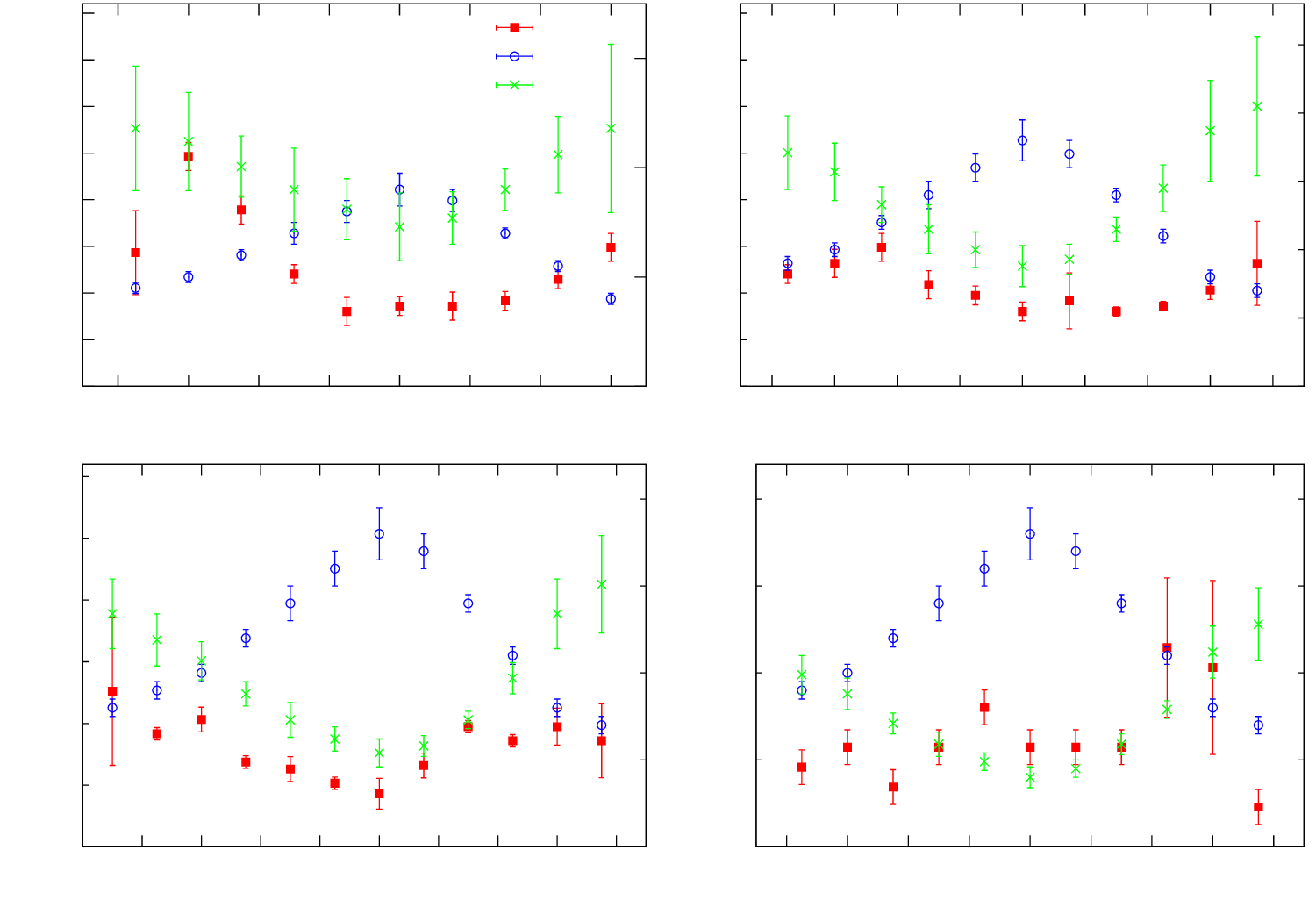}}%
    \gplfronttext
  \end{picture}%
\endgroup

%% file: 0543fig14.tex
\begingroup
  \makeatletter
  \providecommand\color[2][]{%
    \GenericError{(gnuplot) \space\space\space\@spaces}{%
      Package color not loaded in conjunction with
      terminal option `colourtext'%
    }{See the gnuplot documentation for explanation.%
    }{Either use 'blacktext' in gnuplot or load the package
      color.sty in LaTeX.}%
    \renewcommand\color[2][]{}%
  }%
  \providecommand\includegraphics[2][]{%
    \GenericError{(gnuplot) \space\space\space\@spaces}{%
      Package graphicx or graphics not loaded%
    }{See the gnuplot documentation for explanation.%
    }{The gnuplot epslatex terminal needs graphicx.sty or graphics.sty.}%
    \renewcommand\includegraphics[2][]{}%
  }%
  \providecommand\rotatebox[2]{#2}%
  \@ifundefined{ifGPcolor}{%
    \newif\ifGPcolor
    \GPcolorfalse
  }{}%
  \@ifundefined{ifGPblacktext}{%
    \newif\ifGPblacktext
    \GPblacktexttrue
  }{}%
  \let\gplgaddtomacro\g@addto@macro
  \gdef\gplbacktext{}%
  \gdef\gplfronttext{}%
  \makeatother
  \ifGPblacktext
    \def\colorrgb#1{}%
    \def\colorgray#1{}%
  \else
    \ifGPcolor
      \def\colorrgb#1{\color[rgb]{#1}}%
      \def\colorgray#1{\color[gray]{#1}}%
      \expandafter\def\csname LTw\endcsname{\color{white}}%
      \expandafter\def\csname LTb\endcsname{\color{black}}%
      \expandafter\def\csname LTa\endcsname{\color{black}}%
      \expandafter\def\csname LT0\endcsname{\color[rgb]{1,0,0}}%
      \expandafter\def\csname LT1\endcsname{\color[rgb]{0,1,0}}%
      \expandafter\def\csname LT2\endcsname{\color[rgb]{0,0,1}}%
      \expandafter\def\csname LT3\endcsname{\color[rgb]{1,0,1}}%
      \expandafter\def\csname LT4\endcsname{\color[rgb]{0,1,1}}%
      \expandafter\def\csname LT5\endcsname{\color[rgb]{1,1,0}}%
      \expandafter\def\csname LT6\endcsname{\color[rgb]{0,0,0}}%
      \expandafter\def\csname LT7\endcsname{\color[rgb]{1,0.3,0}}%
      \expandafter\def\csname LT8\endcsname{\color[rgb]{0.5,0.5,0.5}}%
    \else
      \def\colorrgb#1{\color{black}}%
      \def\colorgray#1{\color[gray]{#1}}%
      \expandafter\def\csname LTw\endcsname{\color{white}}%
      \expandafter\def\csname LTb\endcsname{\color{black}}%
      \expandafter\def\csname LTa\endcsname{\color{black}}%
      \expandafter\def\csname LT0\endcsname{\color{black}}%
      \expandafter\def\csname LT1\endcsname{\color{black}}%
      \expandafter\def\csname LT2\endcsname{\color{black}}%
      \expandafter\def\csname LT3\endcsname{\color{black}}%
      \expandafter\def\csname LT4\endcsname{\color{black}}%
      \expandafter\def\csname LT5\endcsname{\color{black}}%
      \expandafter\def\csname LT6\endcsname{\color{black}}%
      \expandafter\def\csname LT7\endcsname{\color{black}}%
      \expandafter\def\csname LT8\endcsname{\color{black}}%
    \fi
  \fi
  \setlength{\unitlength}{0.0500bp}%
  \begin{picture}(7200.00,6552.00)%
    \gplgaddtomacro\gplbacktext{%
      \csname LTb\endcsname%
      \put(561,550){\makebox(0,0)[r]{\strut{} 0}}%
      \put(561,1580){\makebox(0,0)[r]{\strut{} 1}}%
      \put(561,2610){\makebox(0,0)[r]{\strut{} 2}}%
      \put(561,3639){\makebox(0,0)[r]{\strut{} 3}}%
      \put(561,4669){\makebox(0,0)[r]{\strut{} 4}}%
      \put(561,5699){\makebox(0,0)[r]{\strut{} 5}}%
      \put(693,330){\makebox(0,0){\strut{} 0}}%
      \put(1783,330){\makebox(0,0){\strut{} 5}}%
      \put(2872,330){\makebox(0,0){\strut{} 10}}%
      \put(3962,330){\makebox(0,0){\strut{} 15}}%
      \put(5052,330){\makebox(0,0){\strut{} 20}}%
      \put(6141,330){\makebox(0,0){\strut{} 25}}%
      \put(110,3485){\rotatebox{90}{\makebox(0,0){$h_{\rm syn}~{[\rm kpc]}$}}}%
      \put(3853,55){\makebox(0,0){$t_{\rm e}~{[\rm 10^6 years]}$}}%
    }%
    \gplgaddtomacro\gplfronttext{%
      \csname LTb\endcsname%
      \put(6488,1203){\makebox(0,0)[r]{\strut{}6.2 cm}}%
      \csname LTb\endcsname%
      \put(6488,967){\makebox(0,0)[r]{\strut{}20 cm}}%
      \csname LTb\endcsname%
      \put(6488,731){\makebox(0,0)[r]{\strut{}90 cm}}%
    }%
    \gplbacktext
    \put(0,0){\includegraphics{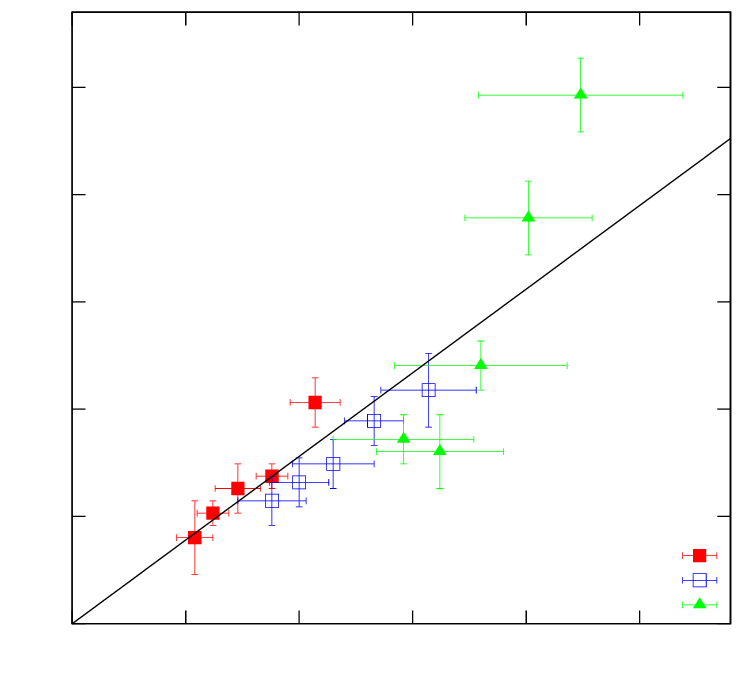}}%
    \gplfronttext
  \end{picture}%
\endgroup

%% file: 0543fig15.tex
\begingroup
  \makeatletter
  \providecommand\color[2][]{%
    \GenericError{(gnuplot) \space\space\space\@spaces}{%
      Package color not loaded in conjunction with
      terminal option `colourtext'%
    }{See the gnuplot documentation for explanation.%
    }{Either use 'blacktext' in gnuplot or load the package
      color.sty in LaTeX.}%
    \renewcommand\color[2][]{}%
  }%
  \providecommand\includegraphics[2][]{%
    \GenericError{(gnuplot) \space\space\space\@spaces}{%
      Package graphicx or graphics not loaded%
    }{See the gnuplot documentation for explanation.%
    }{The gnuplot epslatex terminal needs graphicx.sty or graphics.sty.}%
    \renewcommand\includegraphics[2][]{}%
  }%
  \providecommand\rotatebox[2]{#2}%
  \@ifundefined{ifGPcolor}{%
    \newif\ifGPcolor
    \GPcolorfalse
  }{}%
  \@ifundefined{ifGPblacktext}{%
    \newif\ifGPblacktext
    \GPblacktexttrue
  }{}%
  \let\gplgaddtomacro\g@addto@macro
  \gdef\gplbacktext{}%
  \gdef\gplfronttext{}%
  \makeatother
  \ifGPblacktext
    \def\colorrgb#1{}%
    \def\colorgray#1{}%
  \else
    \ifGPcolor
      \def\colorrgb#1{\color[rgb]{#1}}%
      \def\colorgray#1{\color[gray]{#1}}%
      \expandafter\def\csname LTw\endcsname{\color{white}}%
      \expandafter\def\csname LTb\endcsname{\color{black}}%
      \expandafter\def\csname LTa\endcsname{\color{black}}%
      \expandafter\def\csname LT0\endcsname{\color[rgb]{1,0,0}}%
      \expandafter\def\csname LT1\endcsname{\color[rgb]{0,1,0}}%
      \expandafter\def\csname LT2\endcsname{\color[rgb]{0,0,1}}%
      \expandafter\def\csname LT3\endcsname{\color[rgb]{1,0,1}}%
      \expandafter\def\csname LT4\endcsname{\color[rgb]{0,1,1}}%
      \expandafter\def\csname LT5\endcsname{\color[rgb]{1,1,0}}%
      \expandafter\def\csname LT6\endcsname{\color[rgb]{0,0,0}}%
      \expandafter\def\csname LT7\endcsname{\color[rgb]{1,0.3,0}}%
      \expandafter\def\csname LT8\endcsname{\color[rgb]{0.5,0.5,0.5}}%
    \else
      \def\colorrgb#1{\color{black}}%
      \def\colorgray#1{\color[gray]{#1}}%
      \expandafter\def\csname LTw\endcsname{\color{white}}%
      \expandafter\def\csname LTb\endcsname{\color{black}}%
      \expandafter\def\csname LTa\endcsname{\color{black}}%
      \expandafter\def\csname LT0\endcsname{\color{black}}%
      \expandafter\def\csname LT1\endcsname{\color{black}}%
      \expandafter\def\csname LT2\endcsname{\color{black}}%
      \expandafter\def\csname LT3\endcsname{\color{black}}%
      \expandafter\def\csname LT4\endcsname{\color{black}}%
      \expandafter\def\csname LT5\endcsname{\color{black}}%
      \expandafter\def\csname LT6\endcsname{\color{black}}%
      \expandafter\def\csname LT7\endcsname{\color{black}}%
      \expandafter\def\csname LT8\endcsname{\color{black}}%
    \fi
  \fi
  \setlength{\unitlength}{0.0500bp}%
  \begin{picture}(7200.00,6552.00)%
    \gplgaddtomacro\gplbacktext{%
      \csname LTb\endcsname%
      \put(693,550){\makebox(0,0)[r]{\strut{} 0}}%
      \put(693,1365){\makebox(0,0)[r]{\strut{} 0.5}}%
      \put(693,2181){\makebox(0,0)[r]{\strut{} 1}}%
      \put(693,2996){\makebox(0,0)[r]{\strut{} 1.5}}%
      \put(693,3811){\makebox(0,0)[r]{\strut{} 2}}%
      \put(693,4626){\makebox(0,0)[r]{\strut{} 2.5}}%
      \put(693,5442){\makebox(0,0)[r]{\strut{} 3}}%
      \put(693,6257){\makebox(0,0)[r]{\strut{} 3.5}}%
      \put(825,330){\makebox(0,0){\strut{} 0}}%
      \put(1763,330){\makebox(0,0){\strut{} 5}}%
      \put(2700,330){\makebox(0,0){\strut{} 10}}%
      \put(3638,330){\makebox(0,0){\strut{} 15}}%
      \put(4575,330){\makebox(0,0){\strut{} 20}}%
      \put(5513,330){\makebox(0,0){\strut{} 25}}%
      \put(6450,330){\makebox(0,0){\strut{} 30}}%
      \put(110,3485){\rotatebox{90}{\makebox(0,0){$h_{\rm syn}~{[\rm kpc]}$}}}%
      \put(3919,55){\makebox(0,0){$t_{\rm e}~{[\rm 10^6 years]}$}}%
    }%
    \gplgaddtomacro\gplfronttext{%
      \csname LTb\endcsname%
      \put(6488,1203){\makebox(0,0)[r]{\strut{}6.2 cm}}%
      \csname LTb\endcsname%
      \put(6488,967){\makebox(0,0)[r]{\strut{}20 cm}}%
      \csname LTb\endcsname%
      \put(6488,731){\makebox(0,0)[r]{\strut{}90 cm}}%
    }%
    \gplbacktext
    \put(0,0){\includegraphics{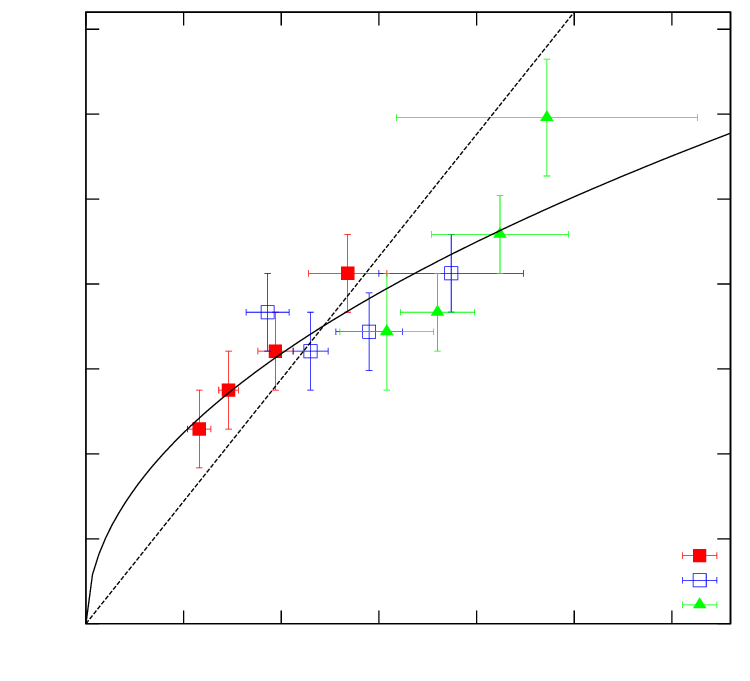}}%
    \gplfronttext
  \end{picture}%
\endgroup

%% file: 0543fig16.tex
\begingroup
  \makeatletter
  \providecommand\color[2][]{%
    \GenericError{(gnuplot) \space\space\space\@spaces}{%
      Package color not loaded in conjunction with
      terminal option `colourtext'%
    }{See the gnuplot documentation for explanation.%
    }{Either use 'blacktext' in gnuplot or load the package
      color.sty in LaTeX.}%
    \renewcommand\color[2][]{}%
  }%
  \providecommand\includegraphics[2][]{%
    \GenericError{(gnuplot) \space\space\space\@spaces}{%
      Package graphicx or graphics not loaded%
    }{See the gnuplot documentation for explanation.%
    }{The gnuplot epslatex terminal needs graphicx.sty or graphics.sty.}%
    \renewcommand\includegraphics[2][]{}%
  }%
  \providecommand\rotatebox[2]{#2}%
  \@ifundefined{ifGPcolor}{%
    \newif\ifGPcolor
    \GPcolorfalse
  }{}%
  \@ifundefined{ifGPblacktext}{%
    \newif\ifGPblacktext
    \GPblacktexttrue
  }{}%
  \let\gplgaddtomacro\g@addto@macro
  \gdef\gplbacktext{}%
  \gdef\gplfronttext{}%
  \makeatother
  \ifGPblacktext
    \def\colorrgb#1{}%
    \def\colorgray#1{}%
  \else
    \ifGPcolor
      \def\colorrgb#1{\color[rgb]{#1}}%
      \def\colorgray#1{\color[gray]{#1}}%
      \expandafter\def\csname LTw\endcsname{\color{white}}%
      \expandafter\def\csname LTb\endcsname{\color{black}}%
      \expandafter\def\csname LTa\endcsname{\color{black}}%
      \expandafter\def\csname LT0\endcsname{\color[rgb]{1,0,0}}%
      \expandafter\def\csname LT1\endcsname{\color[rgb]{0,1,0}}%
      \expandafter\def\csname LT2\endcsname{\color[rgb]{0,0,1}}%
      \expandafter\def\csname LT3\endcsname{\color[rgb]{1,0,1}}%
      \expandafter\def\csname LT4\endcsname{\color[rgb]{0,1,1}}%
      \expandafter\def\csname LT5\endcsname{\color[rgb]{1,1,0}}%
      \expandafter\def\csname LT6\endcsname{\color[rgb]{0,0,0}}%
      \expandafter\def\csname LT7\endcsname{\color[rgb]{1,0.3,0}}%
      \expandafter\def\csname LT8\endcsname{\color[rgb]{0.5,0.5,0.5}}%
    \else
      \def\colorrgb#1{\color{black}}%
      \def\colorgray#1{\color[gray]{#1}}%
      \expandafter\def\csname LTw\endcsname{\color{white}}%
      \expandafter\def\csname LTb\endcsname{\color{black}}%
      \expandafter\def\csname LTa\endcsname{\color{black}}%
      \expandafter\def\csname LT0\endcsname{\color{black}}%
      \expandafter\def\csname LT1\endcsname{\color{black}}%
      \expandafter\def\csname LT2\endcsname{\color{black}}%
      \expandafter\def\csname LT3\endcsname{\color{black}}%
      \expandafter\def\csname LT4\endcsname{\color{black}}%
      \expandafter\def\csname LT5\endcsname{\color{black}}%
      \expandafter\def\csname LT6\endcsname{\color{black}}%
      \expandafter\def\csname LT7\endcsname{\color{black}}%
      \expandafter\def\csname LT8\endcsname{\color{black}}%
    \fi
  \fi
  \setlength{\unitlength}{0.0500bp}%
  \begin{picture}(7200.00,6552.00)%
    \gplgaddtomacro\gplbacktext{%
      \csname LTb\endcsname%
      \put(429,890){\makebox(0,0)[r]{\strut{} 0.1}}%
      \put(429,3796){\makebox(0,0)[r]{\strut{} 1}}%
      \put(3097,220){\makebox(0,0){\strut{} 1}}%
      \put(6726,220){\makebox(0,0){\strut{} 10}}%
      \put(110,3430){\rotatebox{90}{\makebox(0,0){$S~[{\rm Jy}]$}}}%
      \put(3787,165){\makebox(0,0){$\nu~\rm [GHz]$}}%
    }%
    \gplgaddtomacro\gplfronttext{%
      \csname LTb\endcsname%
      \put(6488,6239){\makebox(0,0)[r]{\strut{}northeast}}%
      \csname LTb\endcsname%
      \put(6488,6003){\makebox(0,0)[r]{\strut{}center}}%
      \csname LTb\endcsname%
      \put(6488,5767){\makebox(0,0)[r]{\strut{}southwest}}%
    }%
    \gplbacktext
    \put(0,0){\includegraphics{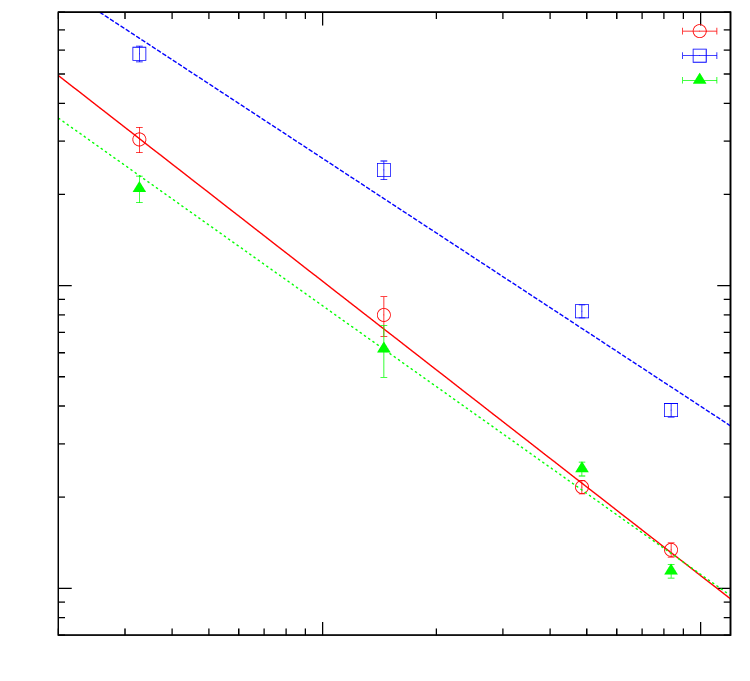}}%
    \gplfronttext
  \end{picture}%
\endgroup

%% file: 0543figA1.tex
\begingroup
  \makeatletter
  \providecommand\color[2][]{%
    \GenericError{(gnuplot) \space\space\space\@spaces}{%
      Package color not loaded in conjunction with
      terminal option `colourtext'%
    }{See the gnuplot documentation for explanation.%
    }{Either use 'blacktext' in gnuplot or load the package
      color.sty in LaTeX.}%
    \renewcommand\color[2][]{}%
  }%
  \providecommand\includegraphics[2][]{%
    \GenericError{(gnuplot) \space\space\space\@spaces}{%
      Package graphicx or graphics not loaded%
    }{See the gnuplot documentation for explanation.%
    }{The gnuplot epslatex terminal needs graphicx.sty or graphics.sty.}%
    \renewcommand\includegraphics[2][]{}%
  }%
  \providecommand\rotatebox[2]{#2}%
  \@ifundefined{ifGPcolor}{%
    \newif\ifGPcolor
    \GPcolorfalse
  }{}%
  \@ifundefined{ifGPblacktext}{%
    \newif\ifGPblacktext
    \GPblacktexttrue
  }{}%
  \let\gplgaddtomacro\g@addto@macro
  \gdef\gplbacktext{}%
  \gdef\gplfronttext{}%
  \makeatother
  \ifGPblacktext
    \def\colorrgb#1{}%
    \def\colorgray#1{}%
  \else
    \ifGPcolor
      \def\colorrgb#1{\color[rgb]{#1}}%
      \def\colorgray#1{\color[gray]{#1}}%
      \expandafter\def\csname LTw\endcsname{\color{white}}%
      \expandafter\def\csname LTb\endcsname{\color{black}}%
      \expandafter\def\csname LTa\endcsname{\color{black}}%
      \expandafter\def\csname LT0\endcsname{\color[rgb]{1,0,0}}%
      \expandafter\def\csname LT1\endcsname{\color[rgb]{0,1,0}}%
      \expandafter\def\csname LT2\endcsname{\color[rgb]{0,0,1}}%
      \expandafter\def\csname LT3\endcsname{\color[rgb]{1,0,1}}%
      \expandafter\def\csname LT4\endcsname{\color[rgb]{0,1,1}}%
      \expandafter\def\csname LT5\endcsname{\color[rgb]{1,1,0}}%
      \expandafter\def\csname LT6\endcsname{\color[rgb]{0,0,0}}%
      \expandafter\def\csname LT7\endcsname{\color[rgb]{1,0.3,0}}%
      \expandafter\def\csname LT8\endcsname{\color[rgb]{0.5,0.5,0.5}}%
    \else
      \def\colorrgb#1{\color{black}}%
      \def\colorgray#1{\color[gray]{#1}}%
      \expandafter\def\csname LTw\endcsname{\color{white}}%
      \expandafter\def\csname LTb\endcsname{\color{black}}%
      \expandafter\def\csname LTa\endcsname{\color{black}}%
      \expandafter\def\csname LT0\endcsname{\color{black}}%
      \expandafter\def\csname LT1\endcsname{\color{black}}%
      \expandafter\def\csname LT2\endcsname{\color{black}}%
      \expandafter\def\csname LT3\endcsname{\color{black}}%
      \expandafter\def\csname LT4\endcsname{\color{black}}%
      \expandafter\def\csname LT5\endcsname{\color{black}}%
      \expandafter\def\csname LT6\endcsname{\color{black}}%
      \expandafter\def\csname LT7\endcsname{\color{black}}%
      \expandafter\def\csname LT8\endcsname{\color{black}}%
    \fi
  \fi
  \setlength{\unitlength}{0.0500bp}%
  \begin{picture}(7200.00,6552.00)%
    \gplgaddtomacro\gplbacktext{%
      \csname LTb\endcsname%
      \put(990,660){\makebox(0,0)[r]{\strut{} 0.5}}%
      \put(990,1683){\makebox(0,0)[r]{\strut{} 0.6}}%
      \put(990,2707){\makebox(0,0)[r]{\strut{} 0.7}}%
      \put(990,3730){\makebox(0,0)[r]{\strut{} 0.8}}%
      \put(990,4753){\makebox(0,0)[r]{\strut{} 0.9}}%
      \put(990,5776){\makebox(0,0)[r]{\strut{} 1}}%
      \put(1122,440){\makebox(0,0){\strut{} 0}}%
      \put(2159,440){\makebox(0,0){\strut{} 0.2}}%
      \put(3196,440){\makebox(0,0){\strut{} 0.4}}%
      \put(4233,440){\makebox(0,0){\strut{} 0.6}}%
      \put(5270,440){\makebox(0,0){\strut{} 0.8}}%
      \put(6307,440){\makebox(0,0){\strut{} 1}}%
      \put(220,3474){\rotatebox{90}{\makebox(0,0){$E_2/E_1$}}}%
      \put(3974,110){\makebox(0,0){$z/h_{\rm B}$}}%
    }%
    \gplgaddtomacro\gplfronttext{%
    }%
    \gplbacktext
    \put(0,0){\includegraphics{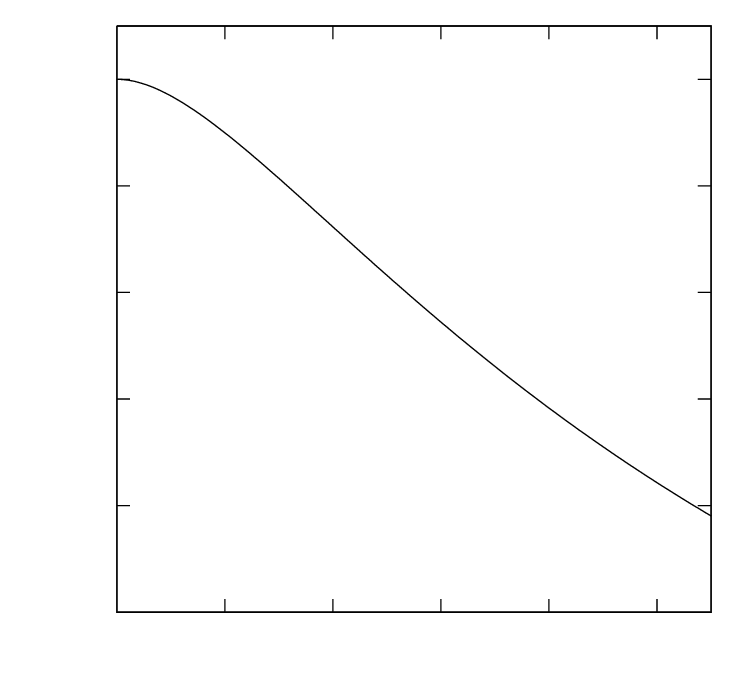}}%
    \gplfronttext
  \end{picture}%
\endgroup